# Physics-Informed Glass-Structure Descriptors for Assessing the Intrinsic Reactivity of Mixed Amorphous–Crystalline Precursors in Alkali-Activated Materials

Zhu Pan[a], Xinru Li[a], Yucheng Wang[a], Samira Hossain [b,c,d], Kai Gong [b,c,d,*]

[a]School of Civil and Transportation Engineering, Hebei University of Technology, Tianjin 300401, China

[b] Department of Civil and Environmental Engineering, Rice University, Houston, Texas 77005, United States

[c] Rice Advanced Materials Institute, Rice University, Houston, Texas 77005, United States

[d] Ken Kennedy Institute, Rice University, Houston, Texas 77005, United States

* Corresponding author. E-mail: kg51@rice.edu

## 1 Abstract

Rapid and reliable assessment of the intrinsic reactivity of amorphous aluminosilicates is critical for their application in alkali-activated materials (AAMs) and blended cements. Although physics-informed glass-structure descriptors have demonstrated strong structure–reactivity relationships for predominantly amorphous systems, their extension to heterogeneous precursors with mixed crystalline-amorphous phases has been limited by the difficulty of isolating amorphous-phase composition. Here, quantitative X-ray diffraction combined with bulk compositional analysis was used to reconstruct the effective amorphous compositions of five fly ashes (FAs) and three ground granulated blast-furnace slags (GGBSs), used as representative heterogeneous aluminosilicate precursors. These compositions served as inputs for molecular dynamics simulations employing a melt-and-quench approach to generate atomic-scale structural models of the glassy phases. Based on these structures, the previously introduced descriptors, i.e., average metal oxygen dissociation energy and average metal oxygen bond strength, were refined to cover a broader compositional space spanning $SiO_2$–$Al_2O_3$–$TiO_2$–$Fe_2O_3$–CaO–MgO–MnO–$Na_2O$–$K_2O$. The refined descriptors exhibit strong inverse correlations with multiple independent reactivity indicators, including cumulative heat release from isothermal calorimetry, bound water content from thermogravimetric analysis, and compressive strength, for both single precursors and binary FA-GGBS blends activated with NaOH. These results demonstrate that physics-informed glass-structure descriptors can be extended from ideal amorphous systems to heterogeneous mixed-phase precursors and capture relative intrinsic reactivity trends in alkaline solutions. The proposed framework provides a transferable, structure-informed basis for comparative assessment of precursor reactivity that complements experimental testing and may inform precursor screening and mix designs for AAM and blended cement systems.

## 2 Introduction

Portland cement (PC) underpins modern infrastructure and is used extensively in bridges, roads, buildings and energy systems. It is the most widely used construction material globally, with an annual production of ~4.1 Gt[1], and demand is projected to increase by 12-23% over 2020-2050[2], driven by population growth and the rehabilitation of aging infrastructure[2]. However, PC clinker production is among the most energy- and carbon-intensive industrial processes, accounting for ~2-3% global energy consumption and 8-9% of anthropogenic $CO_2$ emissions[3]. As a result, decarbonizing cement production is a critical component of global efforts to achieve carbon neutrality by mid-century[2].

One of the most effective strategies for reducing the environmental footprint of cement is the use of blended cement systems, in which a portion of PC clinker is replaced with supplementary cementitious materials (SCMs) with substantially lower associated $CO_2$ emissions[2,4,5]. Among the most widely used SCMs are ground granulated blast-furnace slag (GGBS), a byproduct of steel production with an annual production of ~330 Mt, and fly ash (FA), a residue from coal combustion with a global production of ~700-1100 Mt/year[4]. While nearly all GGBS produced is currently utilized in cementitious applications, less than half of the annually produced FA is utilized[4], with an estimated cumulative reserve of ~3.5 Gt in China and the United States alone[4,6,7]. In addition reducing clinker content, SCM incorporation can be beneficial to many engineering properties of concrete when properly formulated[8], while also mitigating the economic and environmental risks associated with landfilling these industrial byproducts.

Beyond their use in blended systems, GGBS and FA (along with other SCMs) can also serve as precursors in alkali-activated materials (AAMs), a class of clinker-free binders that has attracted increasing attention as an alternative cementitious binders [9–12]. When properly formulated, AAMs can exhibit up to ~40-80% reductions in $CO_2$ emissions relative to PC [13–

[15], while achieving comparable or even superior mechanical, chemical and thermal properties [9,10]. AAMs have also been explored for a range of specialized engineering applications[16,17], including heavy metal immobilization and nuclear waste encapsulation [16,18,19]. Together, the use of GGBS and FA (along with other industrial wastes) in both blended cements and AAM systems exemplifies how industrial byproducts can be transformed from economic and environmental liabilities into value-added resources.

The intrinsic reactivity of SCMs and AAM precursors plays a central role in governing the performance of both blended cements[20] and AAM [9,10] systems. To assess this reactivity, two broad classes of approaches have been developed: experimental testing [21] and empirical modeling[9,22,23]. Experimental methods directly probe reactivity under alkaline conditions and include the strength activity index (as specified in ASTM C311)[24], the Chapelle[25] and Frattini[26] tests, and more recently, calorimetry-based methods such as the $R^3$ test [27,28] and its variants [22,29]. In parallel, a variety of empirical methods based primarily on bulk oxide composition have also been used to assess precursor reactivity, as summarized in references [9,22,23]. Although these composition-based approaches are relatively simple and easy to implement, their predictive reliability and generalizability are limited because they neglect key mineralogical and structural attributes that directly impact dissolution and reaction kinetics.

To address these limitations, physics-based structural descriptors have been developed in recent years to quantify the intrinsic reactivity of amorphous aluminosilicates by explicitly incorporating both chemical composition and atomic-scale structural information, often derived from molecular dynamics (MD) simulations [30–34]. Representative examples include the average metal-oxygen dissociation energy (AMODE)[31], average metal-oxygen bond strength (AMOBS)[32] and topology constraint-based parameters[33,34]. Among these, AMODE provides an overall estimate of the average energy required to break all metal-oxygen bonds within a glassy network, with lower AMODE values corresponding to lower energetic barriers for dissolution and thus higher intrinsic reactivity. These descriptors have demonstrated strong predictive capability across a wide range of amorphous aluminosilicates with compositions representative of slags, fly ashes and volcanic glasses [31,32].

Despite their promise, the applicability of these structural descriptors has largely been restricted to relatively pure, predominantly amorphous systems, such as synthetic glasses or highly vitrified slags. In contrast, many industrial SCMs, particularly fly ashes, are mineralogically heterogeneous and consist of both a reactive amorphous fraction and largely inert crystalline phases [35–38]. The crystalline content of fly ash can vary widely, typically ranging from approximately 40% to 80% [36–38]. This heterogeneity presents a fundamental challenge for applying composition-based reactivity descriptors to real-world materials. Crystalline phases such as quartz ($SiO_2$) and mullite ($Al_6Si_2O_{13}$) contribute minimally to reactivity under alkaline conditions, whereas the amorphous aluminosilicate fraction dominates dissolution and reaction process [18,39]. Consequently, directly applying descriptors developed for fully amorphous systems to heterogeneous SCMs introduces uncertainty and limits their practical applicability.

To address this gap, the present study extends the application of physics-based structural descriptors to SCMs containing mixed amorphous–crystalline phases. Specifically, five FA samples sourced from different coal-fired power plants and three commercially sourced GGBS samples from distinct steel production facilities were investigated. Quantitative X-ray diffraction (QXRD) using an internal standard method was employed to determine the mineralogical compositions of each precursor and to quantify crystalline and amorphous phase fractions. These results were combined with bulk oxide compositions obtained from X-ray fluorescence (XRF) to estimate the chemical composition of the amorphous fraction in each precursor. Force-field MD simulations following a melt-and-quench approach were then used to construct atomic models of the amorphous phases, from which several structural descriptors were calculated, including updated forms of the modified AMODE and AMBOS parameters that depend only on the composition of the amorphous fractions.

The relevance of these descriptors was evaluated by correlating them with experimentally measured reactivity indicators in NaOH-activated systems, including heat release measured by isothermal conduction calorimetry (ICC), bound water content determined by thermogravimetric analysis (TGA), and compressive strength development. Additionally, the phase assemblages and the molecular structure of the resulting AAMs were characterized using XRD and Fourier-transform infrared spectroscopy (FTIR), respectively. Although precursor reactivity was assessed under NaOH-based activation, the descriptors examined here reflect intrinsic properties of the amorphous aluminosilicate network, i.e., metal–oxygen bond energetics, and are therefore expected to capture relative reactivity trends in other alkaline environments where dissolution of the amorphous phase governs reaction progress. By accounting for mineralogical heterogeneity and amorphous-phase chemistry, this integrated approach bridges atomic-scale structural information and macroscopic performance, providing a systematic, structure-informed basis for comparative assessment of heterogeneous SCMs and AAM precursors.

# 3 Materials & Methods

### 3.1.1 Raw materials

Five Class F fly ash (FA1-FA5) samples from different coal-fired power plants in China, along with three ground granulated blast furnace slags (GGBS1-GGBS3) sourced from three integrated iron and steel plants in China were used in this study. For each sample, about 20 kg raw material was supplied and preprocessed using riffle splitters to produce representative samples of specific quantities. Around 200 g of each FA and GGBS sample were collected for detailed characterization, while an additional ~200 g was used to evaluate their reactivity during alkaline activation. The raw materials (i.e., FAs and GGBSs) and the resulting alkali-activated FAs and GGBSs were thoroughly characterized using a range of experimental techniques, as detailed below.

### 3.1.2 Determination of physical properties

The physical characterization of the raw materials consisted of particle size distribution (PSD) analysis and Brunauer–Emmett–Teller (BET) surface area measurements. PSD analysis was conducted using a laser diffraction-based particle size analyzer (PSA, Mastersizer 2000, Malvern Instruments), while BET surface area was measured using nitrogen adsorption at 76.93 K with the ASAP 2010 instrument from Micromeritics. Prior to BET measurement, samples underwent outgassing to eliminate gases and vapors that could potentially be physically adsorbed on particle surfaces [40].

The laser diffraction results, including PSDs and cumulative volume, are presented in Fig. 1, revealing discernible variations among the PSDs of the eight raw materials. Generally, GGBSs displayed narrower size distributions compared to FAs, suggesting a higher degree of size uniformity among GGBS particles. Fig. 1(a) and Table 1 illustrate that GGBS powders exhibit finer characteristics than FA powders, evidenced by smaller peak particle sizes and d10, d50 and d90 values. Notably, FA5 is an exception, featuring a peak particle size of ~7 μm and d10 of ~1.3 μm, which are smaller than those of the GGBSs (~12 μm and ~1.8-2.2 μm).

Interestingly, although FA1-4 exhibit coarser PSDs than the GGBSs based on laser diffraction (Fig. 1(a)), their BET surface areas (1.39-2.19 $m^2/g$, Table 1) are notably higher than those of the GGBSs (0.84-1.06 $m^2/g$). This observation contradicts the anticipated relationship where larger particle sizes are often associated with smaller surface areas. This discrepancy may be attributed to the presence of internal porosity within fly ash particles, a feature that has been widely reported in the literature[41,42]. These internal pores, formed during the rapid cooling and solidification of molten ash, are not captured by laser diffraction but are accessible to nitrogen during BET adsorption analysis. Their size and distribution vary depending on coal source and combustion conditions [43]. Furthermore, fly ash often contains unburned carbon[44,45], which is known to be highly porous and can significantly increase BET surface area relative to the inorganic fraction[45]. This interpretation is supported by the Loss on Ignition (LOI) data in Table 2, which indicates varying residue carbon content among the FA samples. Notably, FA5 exhibits the lowest LOI value among the fly ashes, suggesting minimal residual carbon. This observation aligns with several previous reports that high calcium fly ash generally contain lower levels of unburned carbon compared with low Ca fly ashes [37,46]. Consequently, FA5 displays the smallest BET surface area despite having the lowest $d_{10}$, $d_{50}$ and $d_{90}$ values among all FA samples analyzed (Table 1). Overall, the measured PSDs and BET surface areas for both FAs and GGBSs in this study fall within the range of values previously reported in the literature [47–50].

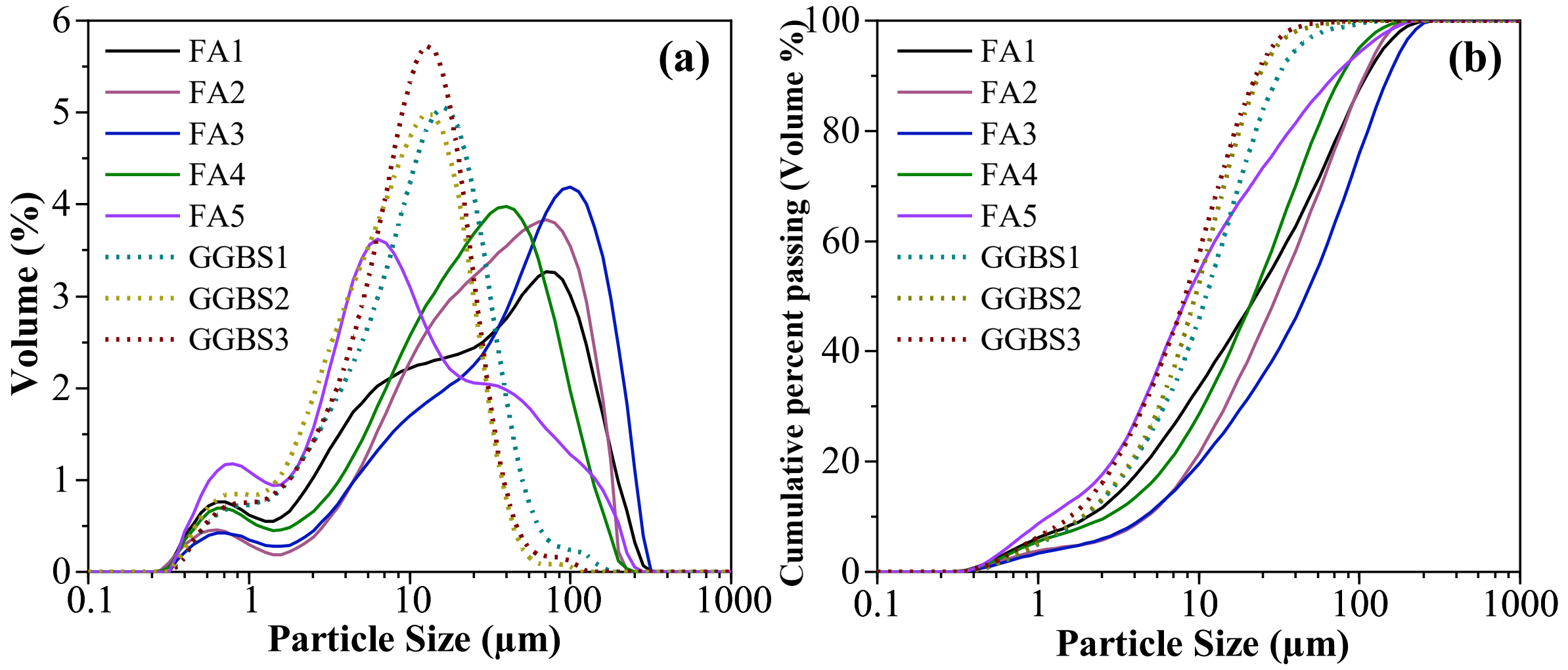

**Fig. 1.** (a) Volume-based particle size distributions and (b) Cumulative volume curves of the five fly ashes (FA) and three ground granulated blast-furnace slags (GGBSs) obtained from laser diffraction-base particle size analysis.

**Table 1.** Summary of particle size parameters ($d_{10}$, $d_{50}$, and $d_{90}$, in μm) for the five fly ashes (FAs) and three ground granulated blast-furnace slags (GGBSs) based on laser diffraction particle size analysis. The BET surface area (in $m^2/g$) of each raw material from $N_2$ adsorption is also given.

| | FA1 | FA2 | FA3 | FA4 | FA5 | GGBS1 | GGBS2 | GGBS3 |
|---|---|---|---|---|---|---|---|---|
| d10 | 2.30 | 5.35 | 5.16 | 2.98 | 1.29 | 2.16 | 2.11 | 1.78 |
| d50 | 25.50 | 34.19 | 51.94 | 25.62 | 9.55 | 12.49 | 10.71 | 9.27 |
| d90 | 124.16 | 120.49 | 170.79 | 86.63 | 78.68 | 35.04 | 26.68 | 25.42 |
| Surface area | 2.19 | 1.66 | 1.68 | 1.39 | 0.83 | 0.84 | 0.88 | 1.06 |

### *3.1.3 Determination of bulk chemical composition and loss on ignition (LOI)*

The bulk chemical compositions of the raw materials were determined using an energy dispersive X-ray fluorescence (XRF) analyzer (ARL QUANT X) following standard procedures [37]. Specifically, the raw materials were ground into a fine powder passing through a 200-mesh sieve. Approximately 2 g of the fine powder was then compressed into a pellet for measurement. The XRF results, summarized in Table 2, indicate that GGBSs primarily consist of CaO, $SiO_2$, MgO, and $Al_2O_3$, with minor amounts of $Fe_2O_3$, MnO, $TiO_2$, $K_2O$, and $Na_2O$, consistent with previous studies[22,35]. In contrast, the fly ashes are predominated composed of $SiO_2$ and $Al_2O_3$, with lesser amounts of $Fe_2O_3$ and CaO and minor contributions from $Na_2O$, $K_2O$, MgO and $TiO_2$, consistent with previous reports[36–38,47]. Notably, the CaO content of FA1–FA5 ranges from 3.0 to 15.9 wt.%, while their combined $SiO_2 + Al_2O_3 + Fe_2O_3$ content falls between ~74 wt.% and ~92 wt.%, satisfying the ASTM C618 chemical criteria for Class F fly ash (CaO<18 wt.% and $SiO_2 + Al_2O_3 + Fe_2O_3 > 50$ wt.%) [51].

For the determination of LOI, each raw material was first oven-dried overnight at 105 °C to remove residual moisture. After recording the initial weight at room temperature, the sample was then ignited for 90 mins at 950 °C in an air ventilated electric muffle furnace. Following cooling, the final sample weight was measured to enable the calculation of the mass loss between 105 °C and 950 °C (i.e., LOI). The measured LOI values, summarized in Table 2, range from 0.5 to 2 wt. %, which is within the limits for Class F fly ash under ASTM C618 [51].

**Table 2.** Bulk chemical composition (wt.%) of the studied fly ashes (FA1-FA5) and ground granulated blast-furnace slags (GGBS1-GGBS3), as measured by X-ray fluorescence (XRF). Loss of ignition (LOI) values at 950 °C are also given.

| Component | FA1 | FA2 | FA3 | FA4 | FA5 | GGBS1 | GGBS2 | GGBS3 |
|---|---|---|---|---|---|---|---|---|
| MgO | 0.709 | 0.399 | 0.648 | 0.51 | 3.06 | 9.08 | 8.04 | 8.02 |
| $Al_2O_3$ | 38.1 | 37.2 | 32 | 37.2 | 17.5 | 13.6 | 15.8 | 15.3 |
| $SiO_2$ | 47.3 | 49.4 | 51.4 | 47.5 | 49.6 | 31.1 | 29 | 31.7 |
| $SO_3$ | 0.916 | 1.42 | 1.88 | 1.8 | 0.882 | 2.42 | 2.92 | 2.35 |
| $K_2O$ | 0.953 | 1.09 | 1.57 | 1.12 | 2.78 | 0.51 | 0.346 | 0.334 |
| CaO | 4.1 | 3.02 | 4.51 | 5.26 | 15.9 | 40.5 | 41.2 | 40.5 |
| $TiO_2$ | 1.57 | 1.46 | 1.33 | 1.51 | 0.668 | 1.38 | 1.21 | 0.565 |
| $Fe_2O_3$ | 4.88 | 4.98 | 5.33 | 3.88 | 7.33 | 0.272 | 0.319 | 0.364 |
| $Na_2O$ | 0.2 | 0.325 | 0.639 | 0.322 | 1.64 | 0.594 | 0.528 | 0.338 |
| $P_2O_5$ | 0.543 | 0.27 | 0.312 | 0.385 | 0.154 | 0.012 | 0.010 | 0.012 |
| MnO | 0.055 | 0 | 0 | 0.052 | 0.124 | 0.352 | 0.351 | 0.27 |
| Others | 0.729 | 0.436 | 0.381 | 0.513 | 0.486 | 0.192 | 0.286 | 0.259 |
| LOI | 1.67 | 0.69 | 0.959 | 1.303 | 0.688 | 0.539 | 1.769 | 0.555 |

### *3.1.4 Determination of mineralogical composition*

The phase compositions of the studied raw materials were determined using Rietveld refinement of quantitative X-ray diffraction (Q-XRD) data, following an internal standard method. This technique is widely reported as a reliable and accurate method for quantifying both crystalline and amorphous phases in complex multiphase systems such as fly ash[36,52–54]. To estimate the amorphous content, 15 wt % of Zincite (ZnO) was used as an internal standard and homogenously mixed with 85 wt % of the raw material (either FA or GGBS). The homogenized mixture was then loaded for XRD analysis using a Bruker D8 Discover diffractometer (Germany). The instrument was operated at 40 kV and 40 mA with Cu K$\alpha$ radiation. XRD data were collected over a 2θ range of 5°and 70° with a step size of 0.02 and a counting time of 1s per step.

Rietveld full-pattern refinements were performed using the TOPAS 4.2 software package[55] (Bruker AXS, 2003–2009). The crystallographic information files (CIF) used for phase identification are summarized in Table S1 of the Supplementary Material, and structural parameters were allowed to vary within ±2 % of their reference values. The broad diffuse scattering arising from the amorphous phase was modeled using a split pseudo-Voigt (SPV) function. To reduce the number of adjustable parameters, the Lorentzian fraction of the SPV function was fixed at 1 on the left and 0.5 on the right, as these values consistently yielded good agreement between calculated and experimental patterns across all samples. The integrated area under the SPV curve was used as the effective scale factor for quantifying the amorphous phase. Diffraction peaks for crystalline phases were fitted using conventional pseudo-Voigt functions while the background was modeled using a first-order polynomial combined with a 1/2θ term, following established practices [52,53].The resulting weighted profile R-factor (Rwp) achieved, a key indicator of fit quality in Rietveld analysis, was below 5% for all samples (Table S3 of the Supplementary Material), indicating excellent agreement between the observed and calculated diffraction patterns.

### *3.2 Modeling the atomic structure of the amorphous phases in FAs and GGBSs*

The chemical compositions of the amorphous components in FAs and GGBSs were estimated by subtracting the oxide contents of the crystalline phases, as quantified via Q-XRD, from the bulk compositions obtained using XRF analysis (Table 2), similar to an early study [44]. With the chemical composition of the amorphous component, the atomic structures of the amorphous components in the raw FAs and GGBSs were generated using force field molecular dynamics (MD) simulations, following the widely adopted "melt-and-quench" approach[56], as similar to our previous MD simulations [31,32,57]. This approach enables the generation of amorphous structures representative of glassy materials. The MD simulations here employed the force field developed by Pedone et al. [58], which is specifically parameterized for silicate-based systems (crystals, melts, and glasses) within the compositional space of $CaO$-$MgO$-$Al_2O_3$-$SiO_2$-$TiO_2$-$FeO$-$Fe_2O_3$-$Na_2O$-$K_2O$. The Pedone force field, as given by Equation (1), consists of three principal interaction terms: (i) a long-range Coulombic term to account for electrostatic interactions between charged species, (ii) a short-range Morse potential that captures bond stretching and dissociation, and (iii) a steep repulsive term that prevents unphysical atomic overlap at very short interatomic distances.

$$U_{ij}(r_{ij}) = \frac{z_i z_j}{r_{ij}} + D_{ij}\left[\left\{1 - e^{-a_{ij}(r_{ij}-r_0)}\right\}^2 - 1\right] + \frac{C_{ij}}{r_{ij}^{12}} \quad (1)$$

Here, $D_{ij}$, $a_{ij}$, $r_0$, and $C_{ij}$ are empirically derived parameters fitted against experimental data for various silicate crystals[58], and are summarized in Table S2 of the Supplementary Material. This force field has been successfully applied to model the atomic structures of silicate glasses [59–61], amorphous GGBS[62,63], and volcanic ashes[32].

Specifically, for each FA and GGBS sample, a simulation cell containing ~24,000 atoms were generated by randomly placing atoms, based on the average amorphous composition, on a uniform grid. This system was first equilibrated at 5000 K for 0.5 ns using the canonical NVT ensemble and the Nosé Hoover thermostat[64] to ensure complete melting and eliminate any memory of the initial configuration. This was followed by a staged cooling process: the structure was first quenched from 5000 K to 2000 K over 1.5 ns, then equilibrated at 2000 K for 1 ns. A second quenching step was then applied, cooling the system from 2000 to 300 K over 1.7 ns, with a subsequent 0.5 ns equilibration at 300 K. These steps were carried out under a canonical NPT ensemble using the Martyna Tobias Klein thermostat and barostat [65]. Finally, an additional 0.5 ns equilibration at 300 K was conducted using the NVT ensemble with the Nosé Hoover thermostat. Throughout the entire melt-quench process, a time step of 1 fs was used.

Structural analysis was performed on 500 structural snapshots extracted from the MD trajectories of the final 0.5 ns of the NVT equilibration at 300 K. This analysis included the calculation of radial distribution functions (RDFs), and the determination of nearest-neighbor interatomic distances and coordination numbers (CNs) of various metal cations within their first coordination shell (details given in Section S3 of Supplementary Material). Previous investigations on CaO-MgO-$Al_2O_3$-$SiO_2$ (CMAS) and CaO-$Al_2O_3$-$SiO_2$ (CAS) glasses[31], as well as volcanic ashes [32], have reported limited variability in interatomic distances and CNs across independent simulations for a given composition. Nevertheless, to ensure statistical robustness, three independent simulations were performed for each amorphous composition. All MD simulations were carried out using the ATK-Forcefield module within the QuantumATK software package [66,67].

### 3.3 ***Characterization of alkali-activated materials (AAMs)***

#### 3.3.1 *Sample preparation*

To assess the alkaline reactivity of the various precursors, each material was activated using a 3 M NaOH solution prepared from 99% reagent-grade NaOH pellets. The optimal activator concentration can vary substantially depending on precursor type (~1-15 M)[48,68–70]; however, to enable direct comparison of reactivity across all precursors, a uniform concentration was employed for this study.

AAM pastes were prepared using a mix proportion of 48.5 g of 3M NaOH solution per 100 g of precursor (FA or GGBS), corresponding to a water-to-binder (w/b) ratio of 0.43. To prepare the binders, the precursor powders were first dry-mixed for 3 mins, followed by an additional 3 mins of mixing during which the NaOH solution was gradually added. The resulting paste was cast into cubic molds (20 mm × 20 mm × 20 mm) and compacted using a vibrating table to minimize entrapped air. Subsequently, the samples were cured at ~23 °C ± 1 °C and relative humidity (RH) above 95% until testing. Subsequent characterization included thermogravimetric analysis (TGA), Fourier transform infrared spectroscopy (FTIR), X-ray diffraction (XRD), and compressive strength tests.

### *3.3.2 Isothermal conduction calorimetry (ICC)*

Isothermal conduction calorimetry (ICC) is a widely used technique for quantifying the reactivity of slag and fly ash by monitoring the heat released during their reaction in alkaline environments [22,27,28,71–76]. ICC measurements were performed on all AAM pastes using an eight-channel TAM Air isothermal calorimeter (TA Instruments), following references [71,72]. Each mixture (48.5g of 3 M NaOH solution for every 100 g of precursor) was hand-mixed for 1 minute followed by 1 minute of mixing on a vortex mixer to ensure homogeneity. Immediately after mixing, 5 g of the paste was transferred into a standard glass ampoule and sealed. The ampoule was then loaded into the calorimeter chamber, alongside a reference ampoule containing 5 g of deionized water. Heat flow data was continuously collected at 23 °C for 7 days, providing both heat flow and cumulative heat data to assess reaction kinetics and extent.

### *3.3.3 Thermogravimetric analysis (TGA)*

TGA and differential thermogravimetric (DTG) analysis were performed on the AAM binders to assess the impact of precursor chemistry on bound water content at the age of 3, 7, and 28 days, thereby providing insights into the extent of reaction [75–78]. Before testing, the samples were ground into fine powder and passed through a 200-mesh sieve. The resulting powders were then immersed in isopropanol for 24 hours to halt hydration by displacing free water from the binder through solvent exchange [79]. The samples were then filtered using a Büchner funnel and vacuum-dried at 40°C for 24 hours. TGA/DTG measurements were carried out using an SDT/Q600 Simultaneous TGA/DSC instrument (TA instruments). For each test, approximately 20 mg of the sample was placed in an alumina crucible and heated from ambient temperature to 1000 °C at a heating rate of 10°C/min under a nitrogen environment purged at a flow rate of 40 ml/min. The recorded weight loss as a function of temperature was used to quantify the evolution of bound water content in each AAM binder over time. To isolate the contributions from the hydration reaction, reference TGA/DTG measurements were also performed on each raw FA and GGBS sample using the same preparation and testing procedures.

### *3.3.4 Fourier transform infrared spectroscopy (FTIR)*

FTIR measurements were performed on each AAM mixture at various activation times over a 28-day period (i.e., 1, 4, 6, 24, 72, 144 and 672 hrs) using a Bruker V8 FTIR instrument. Corresponding measurements were also performed for each raw FA and GGBS sample. For each FTIR measurement, 32 scans were collected from 400 to 4000 $cm^{-1}$ at a resolution of 4 $cm^{-1}$. Prior to each measurement, a background spectrum was collected under the same testing conditions and subsequently subtracted from the sample spectrum. All measurements were conducted under ambient air conditions.

### *3.3.5 Testing of mechanical properties*

50 mm cubic specimens were prepared from each AAM paste mix for compressive strength testing. The prepared specimens were demolded after sealed curing for 24 hrs at room temperature and stored in a standard curing room at 20 ± 2 °C and 95% relative humidity until the designated testing age (1, 3, 7 and 28 days). Compressive strength was measured in accordance with the ASTM C39 using an Instron testing machine at a loading rate of around 0.3 MPa/s. At each testing age, at least three specimens were tested individually, and the average value was reported as the compressive strength.

# 4 Results and Discussion

## 4.1 Characterization of the Amorphous Fraction in the Raw Precursors

Raw precursor materials, especially FA, are known to comprise both crystalline and amorphous phases[35–38,68]. Among these, the amorphous phase, which is often aluminosilicate-rich, is the primary reactive component under alkaline activation [18,39], whereas alumina and silica present in common crystalline phases remain largely inert and do not significantly participate in dissolution or geopolymerization processes[80]. The intrinsic reactivity of these raw precursors, which strongly influence the engineering properties (e.g., strength development) of the resulting AAM[9,10] and blended cement systems[20,47], is therefore closely linked to both the quantity and chemical composition of the amorphous fraction. Accordingly, this section presents a quantitative analysis of the phase assemblages in the raw precursors, followed by chemical and structural characterization of their amorphous components.

### 4.1.1 Phase identification and quantification

Crystalline and amorphous phases in each raw precursor were identified and quantified using XRD combined with Rietveld refinement, employing 15 wt. % zincite as an internal standard. Fig. 2a and 2b show representative XRD patterns and corresponding refinement results for one fly ash sample (FA1) and one slag sample (GGBS1), respectively, along with the simulated diffraction contributions from individual crystalline phases and the amorphous component. Refinement results for the remaining raw precursors are given in Fig. S1 of the Supplementary Material. Good agreement between experimental and calculated diffraction patterns was obtained for all samples, with weighted profile R-factors (Rwp) below 5% (Table S2), confirming the robustness of the refinement procedure. The quantified weight fractions of crystalline and amorphous phases for all FA and GGBS samples are summarized in Table 3. All three GGBS samples are predominantly amorphous (~97-100 wt. %), consistent with previous XRD studies of GGBS[22,35], with only trace crystalline inclusions identified, specifically, gypsum in GGBS1 and quartz in GGBS2.

In contrast, the FA samples exhibit more complex mineralogical profiles, featuring a considerably higher fraction of crystalline content, ranging from ~11 to 47 wt. %. The dominant crystalline phases are mullite and quartz, accompanied by trace amounts of magnetite and gypsum. This mineralogical profile is consistent with prior XRD studies of coal fly ashes [18,36,44,48,81,82]. Correspondingly, the amorphous content of the FA samples varies widely, from ~53 wt.% in FA2 to ~89 wt.% in FA5, highlighting the pronounced heterogeneity among the FA precursors examined here.

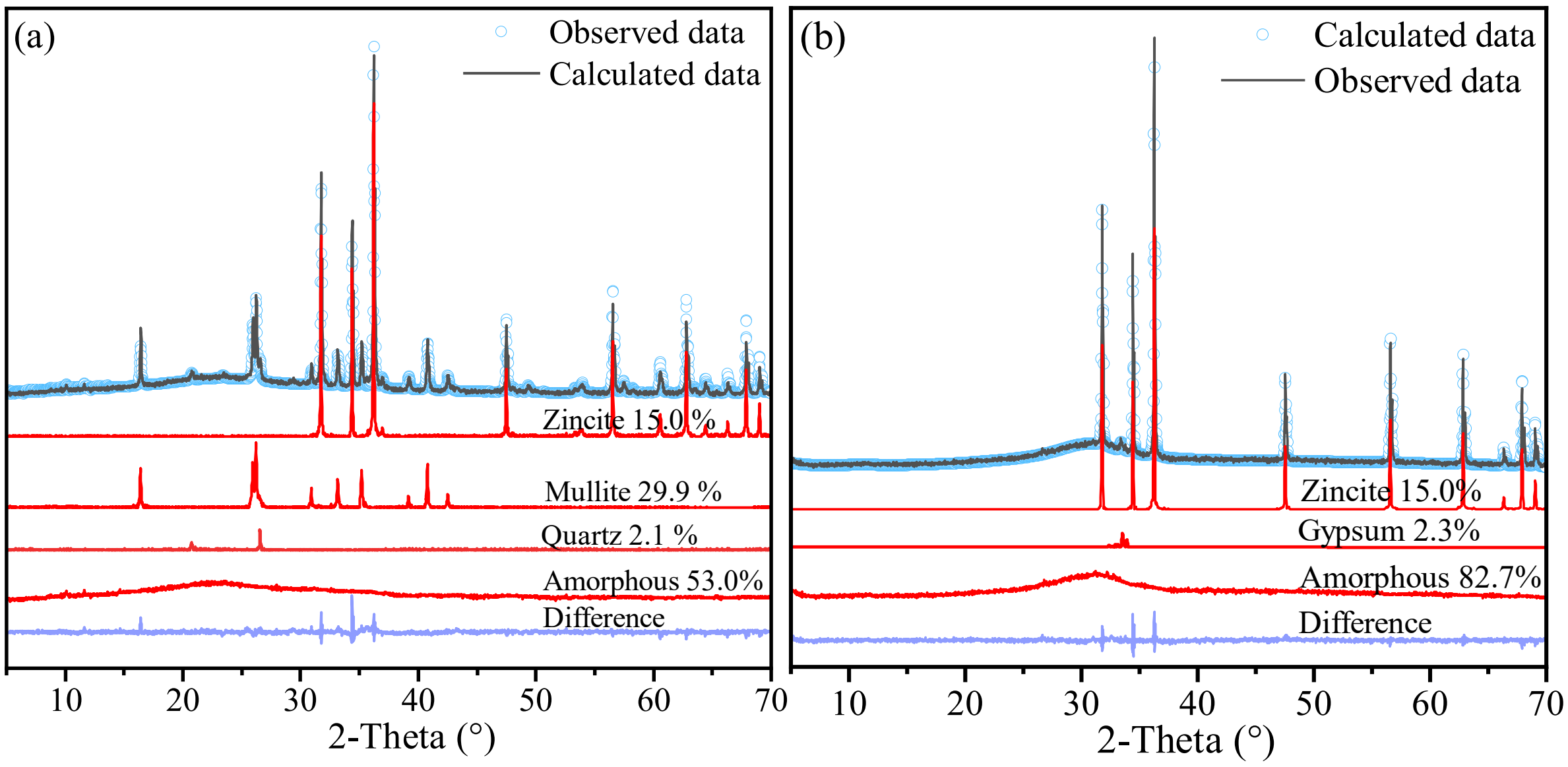

**Fig. 2.** Rietveld refinement of XRD patterns for representative (a) fly ash (FA1) and (b) slag (GGBS1) samples, using 15 wt.% zincite as an internal standard. The measured patterns, calculated fits, and individual contributions from crystalline phases and the amorphous component are shown. The crystallographic models and reference codes used in the refinement, sourced from the Crystallography Open Database, are summarized in Table S1. The corresponding weighted profile R-factors (Rwp) are also given in Table S3. Refinement results for the remaining FA and GGBS samples are provided in Fig. S1 of the Supplementary Material.

**Table 3.** Quantitative phase composition of the five fly ash (FA1–FA5) and three slag (GGBS1–GGBS3) samples, expressed as weight percentages (wt.%), as determined by quantitative X-ray diffraction using Rietveld refinement with an internal standard.

| | FA1 | FA2 | FA3 | FA4 | FA5 | GGBS1 | GGBS2 | GGBS3 |
|---|---|---|---|---|---|---|---|---|
| Mullite ($2.4Al_2O_3 \cdot 1.2SiO_2$) | 32.1 | 39.5 | 18.9 | 25.2 | - | - | - | - |
| Quartz ($SiO_2$) | 3.5 | 7.2 | 4.3 | 1.5 | 10.0 | - | 0.4 | - |
| Magnetite ($Fe_3O_4$) | - | - | - | 0.6 | 1.4 | - | - | - |
| Gypsum ($CaSO_4 \cdot 2H_2O$) | - | - | 0.2 | - | - | 2.8 | - | - |
| Amorphous | 64.4 | 53.2 | 76.7 | 72.7 | 88.6 | 97.2 | 99.6 | 100.0 |

### *4.1.2 Chemical composition of the amorphous component*

The chemical composition of the amorphous component in each raw precursor was estimated by subtracting the oxide contributions of the crystalline, as determined by QXRD (Table 3), from the bulk oxide compositions measured using XRF (Table 2). The resulting oxide compositions of the amorphous fractions, normalized to 100 wt.%, are summarized in Table 4. The estimated amorphous-phase compositions reveal substantial compositional variability among the precursors. Major oxide contents span wide ranges, including $SiO_2$ (27.5-68.2 wt.%), CaO (7.7-48.3 wt. %), $Al_2O_3$ (2.1-16.9 wt.%), $Fe_2O_3$ (0.4-14.4 wt.%), and MgO (0.5-7.6 wt.%). The amorphous phases in the FA samples are predominantly CaO-$Fe_2O_3$-$Al_2O_3$-$SiO_2$ glasses, whereas those in the GGBS samples are dominated by CaO-MgO-$Al_2O_3$-$SiO_2$ glass compositions. Minor constituents, including $Na_2O$, $K_2O$, MnO and $TiO_2$, are also present in varying amounts across the precursors.

**Table 4.** Estimated chemical composition of the amorphous fraction in each fly ash (FA1–FA5) and slag (GGBS1–GGBS3) sample, expressed as oxide weight percentages normalized to 100 wt.%. Compositions were calculated by subtracting crystalline-phase contributions, quantified by QXRD, from bulk oxide compositions measured by XRF.

| Component | FA1 | FA2 | FA3 | FA4 | FA5 | GGBS1 | GGBS2 | GGBS3 |
|---|---|---|---|---|---|---|---|---|
| MgO | 0.9 | 0.6 | 0.7 | 0.5 | 2.8 | 7.6 | 6.7 | 6.6 |
| $Al_2O_3$ | 12.2 | 2.1 | 16.6 | 16.9 | 16.9 | 12.1 | 14.0 | 13.5 |
| $SiO_2$ | 61.3 | 68.2 | 58.7 | 59.5 | 42.2 | 29.8 | 27.5 | 30.2 |
| CaO | 8.5 | 7.7 | 7.8 | 9.8 | 21.4 | 46.6 | 48.3 | 47.4 |
| $Fe_2O_3$ | 11.6 | 14.4 | 10.6 | 7.8 | 10.5 | 0.4 | 0.5 | 0.5 |
| $Na_2O$ | 0.1 | 0.4 | 0.6 | 0.3 | 1.4 | 0.5 | 0.4 | 0.3 |
| $K_2O$ | 1.9 | 2.7 | 2.6 | 2.0 | 3.5 | 0.6 | 0.4 | 0.4 |
| MnO | 0.1 | 0.0 | 0.0 | 0.1 | 0.2 | 0.5 | 0.5 | 0.4 |
| $TiO_2$ | 3.5 | 3.9 | 2.5 | 3.0 | 1.0 | 1.9 | 1.7 | 0.8 |
| Total | 100 | 100 | 100 | 100 | 100 | 100 | 100 | 100 |

*4.1.3 Fourier-transform infrared spectroscopy (FTIR)*

The structural characteristics of the eight raw precursors were further examined using Fourier-transform infrared spectroscopy (FTIR), with spectra in the range of 400-1900 $cm^{-1}$ shown in Fig. 3. The FTIR spectra of the three GGBS samples exhibit largely similar features, reflecting their predominantly amorphous nature (Table 3) and comparable amorphous-phase compositions (Table 4). A prominent absorption band centered around ~960 $cm^{-1}$ is attributed to the asymmetric Si-O-Si/Al stretching vibrations [83,84], while a secondary band near 510 $cm^{-1}$ corresponds to the $\nu_4$(O-Si-O) bending modes of $SiO_4$ tetrahedra[84]. Weak bands observed at ~1410-1460 $cm^{-1}$ and ~875 $cm^{-1}$ indicate the presence of trace carbonate phases[85].

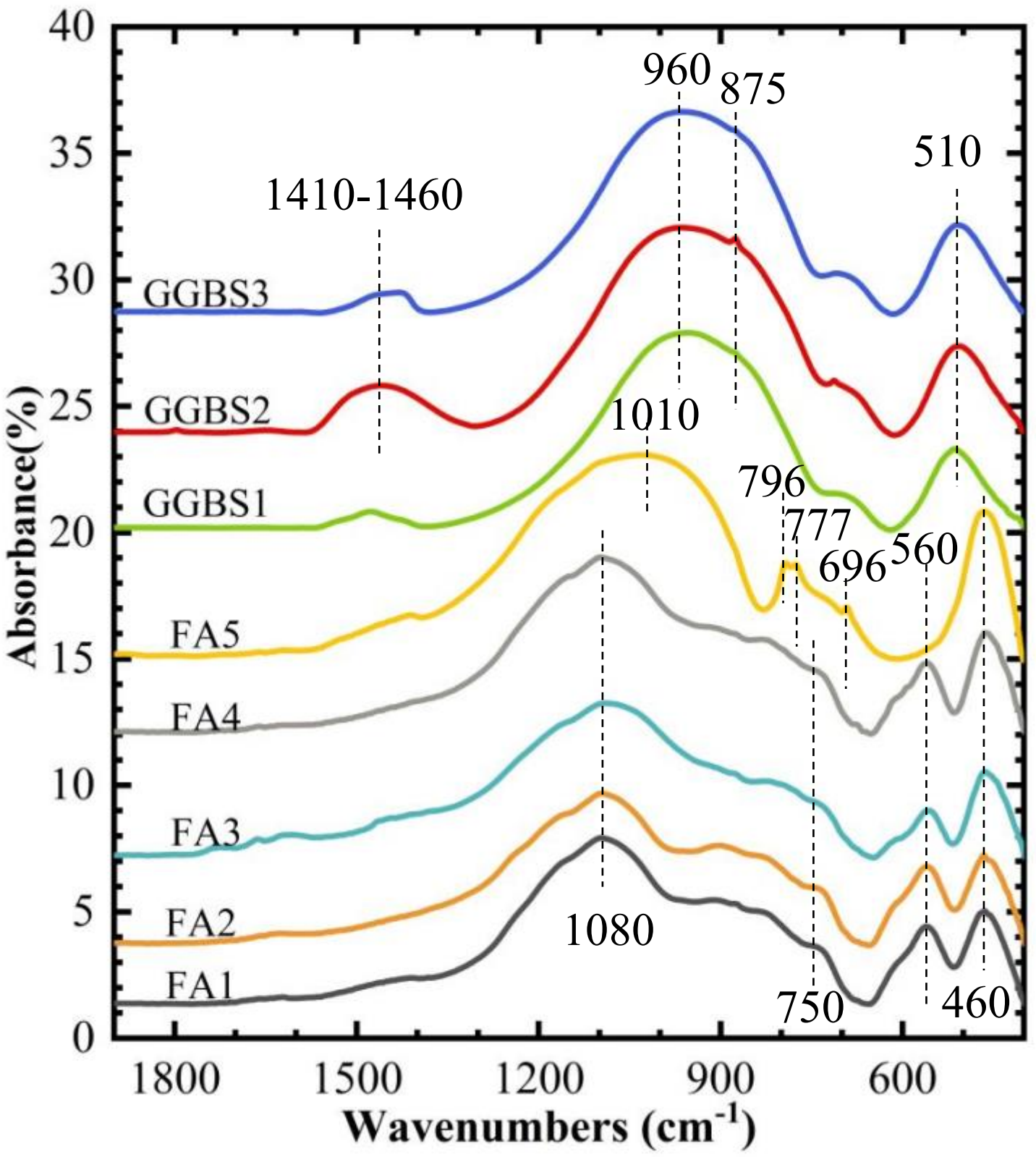

**Fig. 3.** FTIR spectra of raw fly ash (FA1-FA5) and slag (GGBS1-GGBS3) samples, in the range of 400–1900 $cm^{-1}$.

In contrast, the FTIR spectra of the FA samples differ markedly from those of GGBSs, reflecting differences in both chemical and mineralogical compositions (Tables 2-4). The main absorption bands occur near ~1080 $cm^{-1}$ for FA1–FA4 and ~1010 $cm^{-1}$ for FA5. These bands are primarily associated with asymmetric Si-O-Si/Al stretching vibrations within the amorphous aluminosilicate network[83,84], with overlapping contributions from quartz, which exhibits a characteristic band near ~1084 $cm^{-1}$.[85] The position of this main band is often used as an indicator of the degree of network depolymerization, with lower wavenumbers corresponding to more depolymerized structures [83]. Accordingly, the GGBS samples, with the main bands near ~960 $cm^{-1}$, exhibit the most depolymerized amorphous structures, followed by the high-calcium fly ash FA5 (~1010 $cm^{-1}$), and then the low-calcium fly ashes FA1-FA4 (~1080 $cm^{-1}$), which display the lowest degree of depolymerization. This trend is consistent with expectations as the amount of network breakers (e.g., CaO and MgO) that induce depolymerization of the aluminosilicate network[57] increases in the order: FA1-FA4 (8-10 wt.%) < FA5 (23 wt.%) < GGBSs (54-55 wt.%).

Additional features in the FTIR spectra of FA1-FA4 include shoulders at ~1137-1152, ~800-900, and ~750 $cm^{-1}$, along with a peak near 560 $cm^{-1}$, all characteristic of mullite [85,86], the dominant crystalline phase in these samples. These features are absent in FA5, consistent with the QXRD results (Table 3). A band at ~460-470 $cm^{-1}$, attributed to δ(Si-O) bending vibrations in quartz[87], is more pronounced in FA5 than in FA1-FA4, reflecting its higher quartz content (Table 3). Additional quartz-related bands at ~796, ~777, and ~696 $cm^{-1}$ [85,87] are also evident in the FA5 spectrum.

## 4.2 Modeling the Atomic Structure of the Amorphous Component in Raw Precursors

The calculated chemical compositions of the amorphous components in each FA and GGBS sample (Table 4) were used to generate atomistic structural representations using force-field MD simulations, as described in Section 2.2. The number of atoms included in each simulation cell representing the amorphous phase of each precursor is summarized in Table 5. These structural models were generated under the simplifying assumption that the amorphous fraction of each precursor can be represented as a single, homogeneous glassy phase. It is acknowledged that this assumption may not fully capture the complexity of real materials, particularly fly ashes, where multiple glassy phases with distinct chemical compositions may coexist within the amorphous fraction[88]. Nevertheless, this approach provides a consistent and tractable framework for isolating the effects of average amorphous-phase composition on atomic-scale structure and derived descriptors.

Table 5. Number of atoms in the simulation box used to generate atomistic structural models representing the amorphous component in each FA and GGBS sample.

| Component | FA1 | FA2 | FA3 | FA4 | FA5 | GGBS1 | GGBS 2 | GGBS 3 |
|---|---|---|---|---|---|---|---|---|
| MgO | 112 | 76 | 92 | 72 | 380 | 6 | 944 | 920 |
| $Al_2O_3$ | 628 | 108 | 851 | 860 | 908 | 668 | 780 | 744 |
| $SiO_2$ | 5364 | 6036 | 5136 | 5148 | 3836 | 2796 | 2584 | 2828 |
| CaO | 796 | 728 | 730 | 908 | 2088 | 4684 | 4876 | 4752 |
| $Fe_2O_3$ | 380 | 476 | 347 | 252 | 356 | 12 | 16 | 20 |
| $Na_2O$ | 8 | 36 | 50 | 24 | 124 | 44 | 40 | 24 |
| $K_2O$ | 108 | 152 | 147 | 112 | 204 | 32 | 24 | 20 |
| MnO | 8 | 0 | 0 | 8 | 16 | 40 | 40 | 32 |
| $TiO_2$ | 228 | 260 | 165 | 200 | 68 | 136 | 120 | 56 |
| Total number of atoms | 23996 | 23980 | 24128 | 23988 | 23984 | 23984 | 24004 | 24012 |

### *4.2.1 Atomic structural model*

Fig. 4a shows a representative atomistic structural model of the amorphous phase in FA1, illustrating a highly disordered aluminosilicate network. The corresponding total radial distribution function (RDF), shown in Fig. 4b, exhibits well-defined short-range (< ~3 Å) and medium-range (~3-6 Å) order, with no discernible long-range order beyond ~6 Å. These features are characteristic of glassy materials and are consistent with previously reported X-ray and neutron total scattering measurements [31,35,57,71,89–92] as well as prior MD simulations [31,32,71,93,94] on silicate-based glasses. To facilitate peak assignments and local structural analysis, partial RDFs were calculated for individual metal-oxygen (M-O) pairs, where M = Ca, Mg, Al, Si, Ti, Fe, Na, K, and Mn, and are shown in Fig. 4c. The positions of the first peaks in these partial RDF correspond to nearest-neighbor M-O bond distances, which are summarized in Table 6 for all modeled glass compositions. The calculated M-O bond distances (1.59-1.61 Å for Si-O, 1.72-1.73 Å for Al-O, 2.34-2.38 Å for Ca-O, 2.00-2.02 Å for Mg-O, 2.32-2.40 Å for Na-O, 2.65-2.77 Å for K-O, 1.86-1.90 Å for Fe-O, and 1.81-1.84 Å for Ti-O) are in good agreement with experimental and computational values reported for silicate-based glasses, as summarized in ref. [32,57]. The Mn-O bond distances, estimated at 2.09-2.16 Å, also align with values reported from extended X-ray absorption fine structure (EXAFS) analyses of Mn-containing silicate glasses (~2.07-2.17 Å)[95,96]. Across the range of modeled compositions, variations in M–O bond distances are relatively small (typically <5%), indicating that bulk compositional differences exert only a limited influence on local bond lengths within the amorphous network.

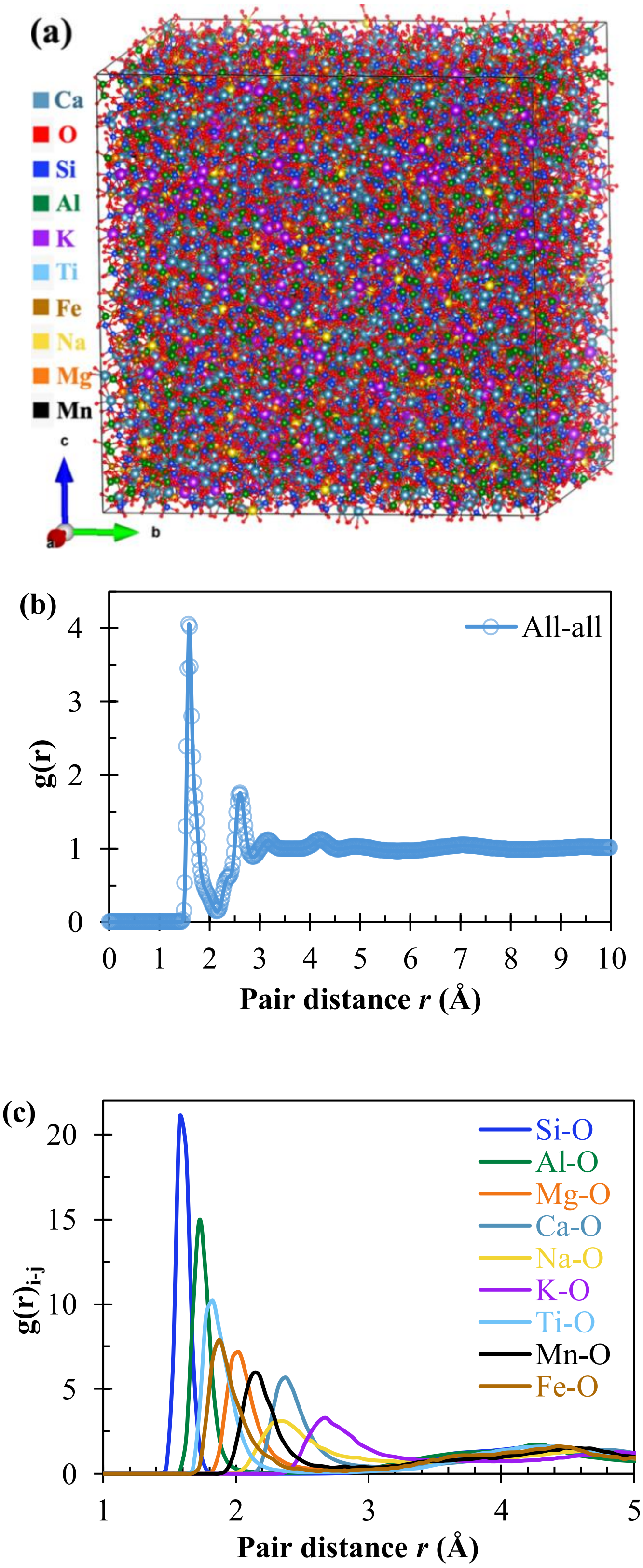

Fig. 4. (a) Representative atomistic structural model of the amorphous component in FA5, (b) corresponding total radial distribution function (RDF), and (c) partial RDFs for individual metal-oxygen (M-O) pairs (M = Ca, Mg, Al, Si, Ti, $^{III}Fe$, Na, K, and Mn) in the glass in (a). All RDFs were computed from the final 500 ps of MD trajectories during the NVT equilibration at 300 K. Calculation details are provided in Section S3 of the Supplementary Material.

Table 6. Nearest-neighbor metal–oxygen (M–O) interatomic distances (Å) for the amorphous components of each FA and GGBS sample, extracted from partial RDFs. Reported values are means with standard deviations (Stdev.) based on three independent MD simulations.

| **Bond distance** | | **FA1** | **FA2** | **FA3** | **FA4** | **FA5** | **GGBS1** | **GGBS2** | **GGBS3** |
|---|---|---|---|---|---|---|---|---|---|
| Si-O | Average | 1.61 | 1.61 | 1.60 | 1.59 | 1.59 | 1.59 | 1.59 | 1.59 |
| | Stdev. | 0.02 | 0.00 | 0.01 | 0.00 | 0.00 | 0.00 | 0.00 | 0.00 |
| Al-O | Average | 1.73 | 1.73 | 1.72 | 1.73 | 1.73 | 1.73 | 1.73 | 1.73 |
| | Stdev. | 0.00 | 0.02 | 0.01 | 0.00 | 0.00 | 0.00 | 0.00 | 0.00 |
| Ca-O | Average | 2.38 | 2.34 | 2.38 | 2.37 | 2.36 | 2.35 | 2.35 | 2.36 |
| | Stdev. | 0.01 | 0.02 | 0.02 | 0.02 | 0.01 | 0.01 | 0.00 | 0.01 |
| Mg-O | Average | 2.01 | 2.01 | 2.02 | 2.01 | 2.00 | 2.01 | 2.01 | 2.01 |
| | Stdev. | 0.04 | 0.02 | 0.01 | 0.00 | 0.01 | 0.00 | 0.00 | 0.00 |
| Na-O | Average | 2.40 | 2.36 | 2.37 | 2.40 | 2.36 | 2.32 | 2.36 | 2.37 |
| | Stdev. | 0.04 | 0.08 | 0.05 | 0.06 | 0.01 | 0.02 | 0.06 | 0.05 |
| K-O | Average | 2.73 | 2.72 | 2.77 | 2.77 | 2.69 | 2.68 | 2.65 | 2.70 |
| | Stdev. | 0.02 | 0.01 | 0.07 | 0.02 | 0.05 | 0.03 | 0.07 | 0.06 |
| $^{III}Fe$-O | Average | 1.87 | 1.86 | 1.87 | 1.87 | 1.88 | 1.88 | 1.88 | 1.90 |
| | Stdev. | 0.00 | 0.01 | 0.02 | 0.02 | 0.01 | 0.01 | 0.02 | 0.05 |
| Mn-O | Average | 2.09 | NA | 2.14 | 2.12 | 2.16 | 2.16 | 2.12 | 2.14 |
| | Stdev. | 0.09 | NA | 0.08 | 0.06 | 0.06 | 0.03 | 0.02 | 0.02 |
| Ti-O | Average | 1.78 | 1.79 | 1.80 | 1.80 | 1.84 | 1.81 | 1.82 | 1.84 |
| | Stdev. | 0.01 | 0.00 | 0.01 | 0.01 | 0.05 | 0.02 | 0.01 | 0.03 |

The average oxygen coordination numbers (CNs), defined as the number of oxygen atoms within the first coordination shell of each metal cation, were calculated using fixed cutoff distances specific to each M-O pair (details provided in Section S3 of the Supplementary Material). These cutoff values were determined from the first minima in the corresponding partial RDFs, as commonly adopted in the literature[32,57,94] and were held constant across all samples to facilitate direct comparison. The applied cutoff values were 2.0 Å for Si–O, 2.3 Å for Al–O, 2.7 Å for Mg–O, 3.1 Å for Ca–O, 3.4 Å for Na–O, 3.8 Å for K–O, 2.6 Å for $^{III}$Fe–O, 3.0 Å for Mn–O, and 2.6 Å for Ti–O. The resulting average CNs are summarized in Table 7 and are consistent with experimental and computational data reported on silicate-based glasses, as summarized in ref. [32,57]. Notably, Mn–O coordination numbers range from ~5.1 to ~5.9 across the modeled glass compositions, in good agreement with values reported from EXAFS studies of Mn-containing silicate glasses (~4.4-5.9)[95,96].

Table 7. Average oxygen coordination numbers (CNs) for individual metal cations in the amorphous phase of each FA and GGBS sample. Reported values are means with standard deviations (Stdev.) based on three independent MD production runs.

| M-O pair | | FA1 | FA2 | FA3 | FA4 | FA5 | GGBS1 | GGBS2 | GGBS3 |
|---|---|---|---|---|---|---|---|---|---|
| Si-O | Average | 4.00 | 4.00 | 4.00 | 4.00 | 4.00 | 4.00 | 4.00 | 4.00 |
| | Stdev. | 0.00 | 0.00 | 0.00 | 0.00 | 0.00 | 0.00 | 0.00 | 0.00 |
| Al-O | Average | 4.10 | 4.03 | 4.12 | 4.11 | 4.11 | 4.09 | 4.07 | 4.12 |
| | Stdev. | 0.03 | 0.03 | 0.01 | 0.04 | 0.01 | 0.05 | 0.04 | 0.03 |
| Ca-O | Average | 6.65 | 6.34 | 6.70 | 6.64 | 6.55 | 6.43 | 6.40 | 6.47 |
| | Stdev. | 0.08 | 0.06 | 0.12 | 0.03 | 0.08 | 0.01 | 0.08 | 0.02 |
| Mg-O | Average | 4.67 | 4.60 | 4.59 | 4.81 | 4.78 | 4.74 | 4.76 | 4.76 |
| | Stdev. | 0.17 | 0.10 | 0.03 | 0.07 | 0.11 | 0.02 | 0.02 | 0.05 |
| Na-O | Average | 8.67 | 7.92 | 8.25 | 8.33 | 7.74 | 7.37 | 7.32 | 7.37 |
| | Stdev. | 0.33 | 0.47 | 0.55 | 0.50 | 0.08 | 0.30 | 0.78 | 0.36 |
| K-O | Average | 10.82 | 10.41 | 10.70 | 10.83 | 10.30 | 9.36 | 9.70 | 9.59 |
| | Stdev. | 0.14 | 0.13 | 0.29 | 0.19 | 0.18 | 0.15 | 0.55 | 0.19 |
| Fe-O | Average | 4.50 | 4.36 | 4.49 | 4.57 | 4.69 | 4.59 | 4.93 | 4.64 |
| | Stdev. | 0.02 | 0.05 | 0.09 | 0.06 | 0.03 | 0.10 | 0.12 | 0.30 |
| Mn-O | Average | 5.49 | NA | 5.92 | 5.84 | 5.07 | 5.79 | 5.30 | 5.51 |
| | Stdev. | 0.50 | NA | 0.15 | 0.12 | 1.69 | 0.39 | 0.19 | 0.31 |
| Ti-O | Average | 4.46 | 4.47 | 4.61 | 4.71 | 5.02 | 5.04 | 5.01 | 5.37 |
| | Stdev. | 0.11 | 0.10 | 0.20 | 0.12 | 0.30 | 0.18 | 0.12 | 0.12 |

### *4.2.2 Calculation of structural descriptors*

Based on the modeled glass compositions (Table 5) and their corresponding MD-derived structural information (Tables 6 and 7), two physics-based structural descriptors, previously developed in our earlier work [31,32], were calculated to assess intrinsic reactivity of aluminosilicate glasses. The first descriptor, the average metal-oxygen bond strength (AMOBS), is derived from the classical bond valence model and is calculated using Equations (2) and (3):

$$AMOBS = \frac{\sum N_M \cdot CN_M \cdot s_M}{\sum N_M} \quad (2)$$

$$s_M = e^{(R_{0_{M-O}} - R_{M-O})/b} \quad (3)$$

where, $N_M$ represents the number of metal cations of type $M$ (where $M$ = Si, Al, Ti, $^{III}$Fe, Mg, Ca, Na, K and Mn; see Table 5), $CN_M$ is the average coordination number (Table 7), and $s_M$ is the bond valence reflecting the relative strength of individual M-O bonds[97]. The bond valence is expressed as a function of the average M-O bond length, $R_{M-O}$ (Table 6), and an empirical reference bond length, $R_{0_{M-O}}$, specific to each M-O pair, which are given in Table S4 of Supplementary Material (sourced from ref. [97]). The exponential term uses a universal empirical constant $b$ set to 0.37.[98]

The second descriptor, the average metal-oxygen dissociation energy (AMODE), was originally introduced in ref. [31] to assess the intrinsic reactivity of CaO–(MgO–)$Al_2O_3$–$SiO_2$ C(M)AS glasses in alkaline environments, and later extended to volcanic glasses [32], as defined by Equation (4):

$$AMODE = \frac{\sum N_M \cdot CN_M \cdot BS_{(M-O)}}{\sum N_M} \quad (4)$$

where $N_M$, $CN_M$ and $BS_{(M-O)}$ denote the number of cations (Table 5), coordination number (Table 7) and average energy required to break a single M-O bond, respectively. The $BS_{M-O}$ values for different M-O pair were obtained from reference [99] and are given in Table S5 of the Supplementary Material.

Because the coordination number $CN_M$ vary only modestly (within ±7%) for a given cation across the modeled compositions, simplified formulations were calculated following prior work [100]. Specifically, fixed average coordination numbers were calculated for each cation type by averaging CN values across all modeled glasses. This simplification enables the calculation of composition-only forms of the descriptors, referred to here as modified AMOBS and modified AMODE, defined in Equations (5) and (6).

$$Modified\ AMOBS = \frac{4.30N_{Si}+3.32N_{Al}+4.92N_{Ti}+3.32N_{Fe}+2.22N_{Ca}+1.99N_{Mg}+2.22N_{Mn}+1.70N_{Na}+2.12N_K}{N_{Si}+N_{Al}+N_{Ti}+N_{Fe}+N_{Ca}+N_{Mg}+N_{Mn}+N_{Na}+N_K} \quad (5)$$

$$Modified\ AMODE = \frac{424N_{Si}+368N_{Al}+353N_{Ti}+331N_{Fe}+209N_{Ca}+174N_{Mg}+198N_{Mn}+157N_{Na}+133N_K}{N_{Si}+N_{Al}+N_{Ti}+N_{Fe}+N_{Ca}+N_{Mg}+N_{Mn}+N_{Na}+N_K} \quad (6)$$

The calculated AMOBS and AMODE values, together with their modified forms, are compared in Fig. 5. The close agreement between the original and modified descriptors (Figs. 5a and 5b) indicates that the use of fixed coordination numbers is a reasonable approximation. Moreover, the modified AMOBS and AMODE exhibit a strong linear correlation ($R^2 = 1.00$; Fig. 5c), despite being derived from different theoretical frameworks. Both descriptors serve as proxies for the energetic stability of M-O network, with higher values corresponding to stronger bonding and, consequently, lower intrinsic chemical reactivity. The ability of these descriptors to capture differences in precursor reactivity is evaluated in Section 4.3.

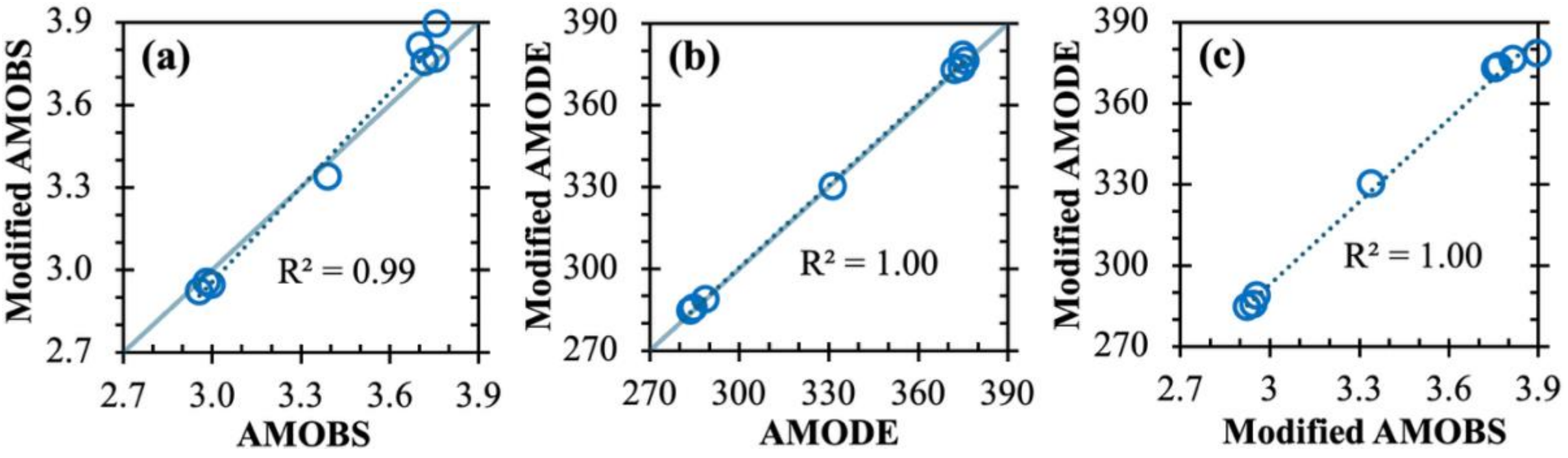

Fig. 5. Comparison of (a) AMOBS and modified AMOBS, (b) AMODE and modified AMODE, and (c) modified AMOBS and modified AMODE. Solid lines in (a) and (b) denote the line of equality, while dotted lines represent linear regression fits. Corresponding $R^2$ values are shown in each figure.

## 4.3 Characterization of Alkaline Reactivity of FA and GGBS samples

This section evaluates the relative alkaline reactivity of the five FA and three GGBS samples using complementary experimental techniques, including isothermal conduction calorimetry (ICC), thermogravimetric analysis (TGA) and compressive strength testing. In addition, the reaction products formed during alkaline activation are characterized using FTIR and XRD.

### *4.3.1 Isothermal conduction calorimetry (ICC)*

ICC is a widely used technique for probing reaction kinetics and heat evolution in alkali-activated materials [72–76,101,102]. Here, ICC was employed to evaluate the reactivity of the eight precursors activated with 3 M NaOH solution. The normalized heat flow and cumulative heat release over a period of ~7 days are displayed in Fig. 6a and 6b, respectively. All samples exhibit a sharp initial heat flow peak within ~0.5 hours after mixing. Such early peaks are commonly observed in both AAM[72,101,102] and OPC-based systems[103], and have been attributed primarily to wetting and rapid initial dissolution upon contact with the solution[101–103], although a recent study suggests that surface wetting dominates this response in alkali-activated slags[72]. For the GGBS samples, a second, broader exothermic peak appears approximately 2-4 hour after mixing, consistent with previous ICC studies on similar NaOH-activated slag system[72]. Among the three GGBS samples, this second peak is slightly delayed for GGBS3, suggesting slower early-age reaction kinetics. However, Fig. 6b shows that GGBS3 exhibits a higher cumulative heat release after ~10 hrs, implying a more sustained reaction over time. Such discrepancy has been reported in prior studies on AAS, where faster early-age reaction at higher alkaline concentration can be accompanied with lower cumulative heat release at later ages[102].

In contrast to GGBS, the FA-based systems exhibit substantially lower heat flows throughout the measurement period. Among the FA samples, the high-calcium FA5 displays a distinct secondary reaction peak at ~3 hrs, whereas FA1–FA4, which have lower calcium contents, exhibit only weak shoulders with significantly lower intensities. These trends are also reflected in the cumulative heat release curves (Fig. 6b), suggesting that GGBS samples are the most reactive, followed by FA5 and then FA1-4. Cumulative heat release values at 1, 3 and 7 days are summarized in Table 8. Subtle differences in reactivity are evident among the GGBS samples and among FA1–FA4, likely reflecting combined effects of particle size, mineralogical composition, and intrinsic chemical reactivity of the amorphous phase.

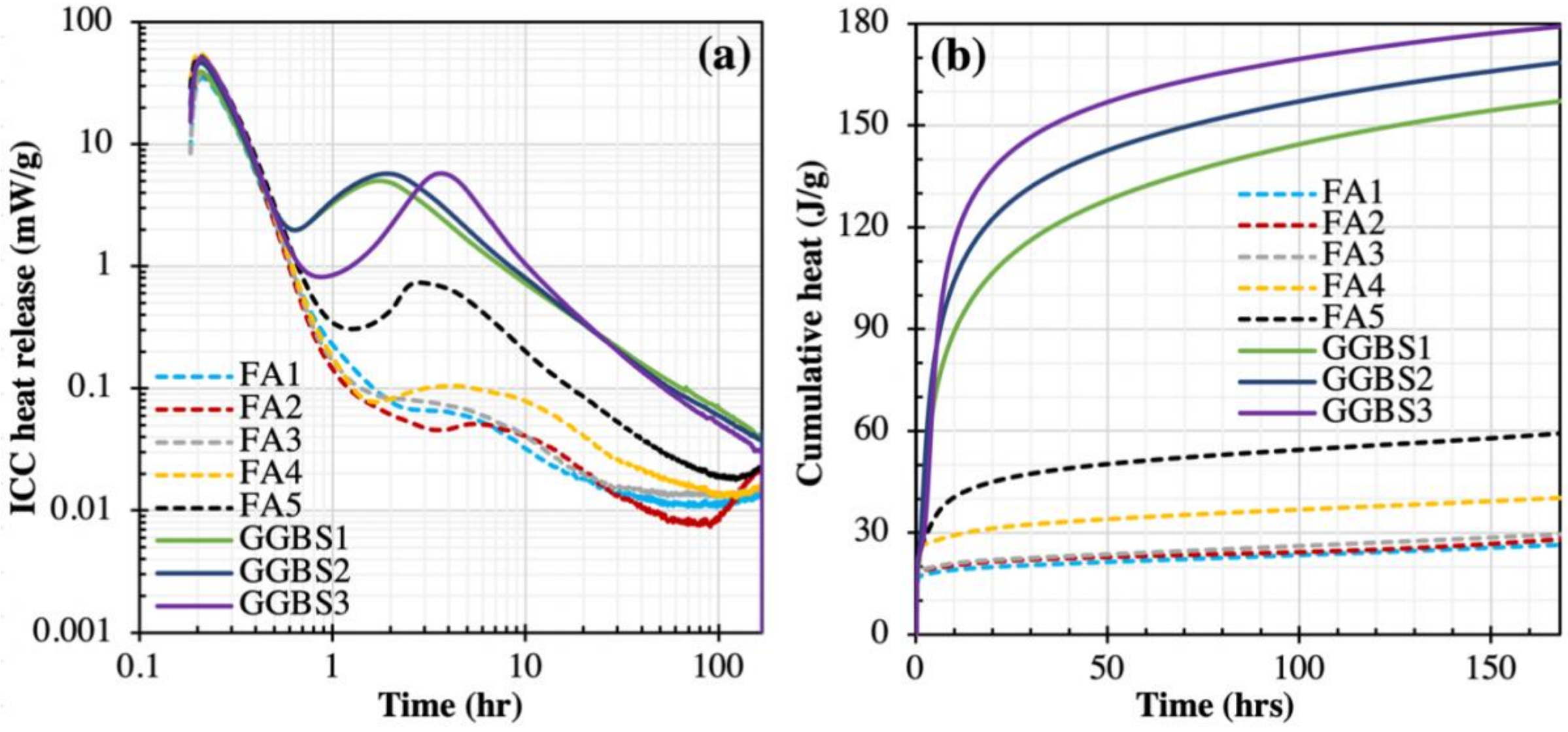

**Fig. 6.** (a) Heat flow (mW/g) and (b) cumulative heat release (J/g) measured by isothermal conduction calorimetry (ICC) for NaOH-activated fly ash (FA1-FA5) and slag (GGBS1-GGBS3) samples.

Table 8. Cumulative heat release (J/g) at 1, 3 and 7 days obtained from ICC measurements for NaOH-activated fly ash (FA1-5) and slag (GGBS1-3) samples.

| | Precursor for the AAM | | | | | | | |
|---|---|---|---|---|---|---|---|---|
| | FA1 | FA2 | FA3 | FA4 | FA5 | GGBS1 | GGBS2 | GGBS3 |
| 1 | 20.15 | 21.73 | 22.24 | 31.83 | 46.12 | 110.95 | 126.9 | 141.6 |
| 3 | 22.29 | 23.48 | 24.72 | 35.35 | 52.34 | 136.59 | 150.24 | 163.72 |
| 7 | 26.62 | 27.77 | 29.32 | 40 | 58.94 | 156.56 | 168 | 178.74 |

### *4.3.2 TGA bound water content and reaction products*

Fig. 7 shows the TGA curves (solid lines) and corresponding derivative thermogravimetric (DTG) curves (dotted lines) for representative NaOH-activated slag (GGBS1) and fly ash (FA1) samples after 3, 7 and 28 days of curing. For reference, the TGA and DTG profiles of the unreacted raw (FA1 and GGBS1) are also included. The corresponding TGA data for FA2-FA5 and GGBS2-GGBS3 is provided in Fig. S2 of Supplementary Material. For NaOH-activated GGBS1, the DTG(TGA) curves (Fig. 7a) exhibit a pronounced peak (weight loss) between 50 °C and 200 °C, which grows with curing age. Because all samples underwent isopropanol exchange prior to TGA testing to remove free water, this mass loss is primarily attributed to dehydration of chemically bound water associated with C-(N)-A-S-H-type gel formation [75,76], with minor contributions from physically absorbed water. The formation of C-(N)-A-S-H-type gel is confirmed by FTIR spectra (Fig. 8a and Fig. S3 in the Supplementary Materials), , where characteristic bands near ~950 $cm^{-1}$, corresponding to asymmetric Si–O–T (T = Si or Al) stretching vibrations,[104] emerge as early as 1 hr and increase in intensity with curing time. This rapid gel formation aligns with prior reports for NaOH-activated slag [71,72,102]. This interpretation is further supported by the 28-day XRD patterns (Fig. 9), which show features (2θ at ~7 and ~29°) characteristic of poorly crystalline C-S-H-type gel [75,76].

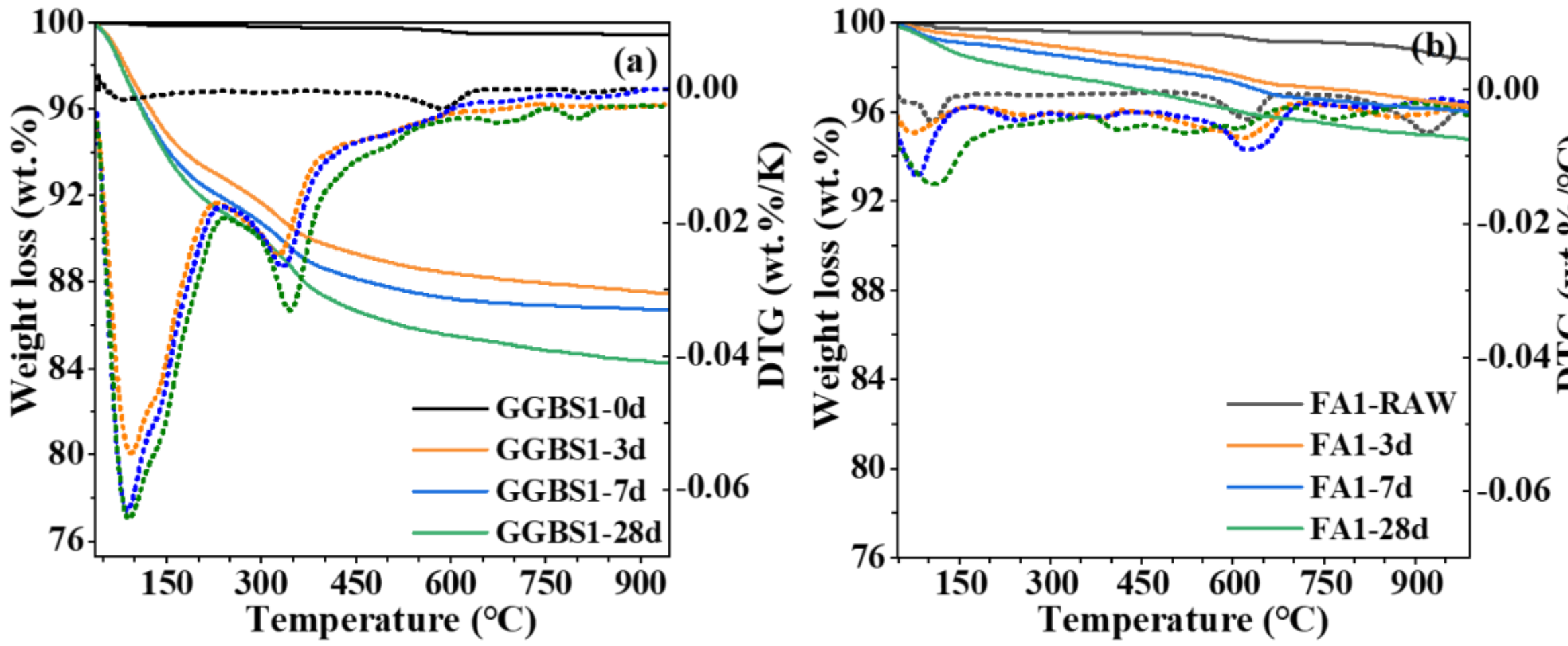

**Fig. 7.** Thermogravimetric analysis (TGA, solid lines) and derivative thermogravimetric (DTG, dotted lines) curves for NaOH-activated (a) slag (GGBS1) and (b) fly ash (FA1) samples after 3, 7, and 28 days of curing. Corresponding TGA/DTG curves of the unreacted raw precursors are included for comparison.

A broader secondary DTG peak observed between approximately 250 and 400 °C (Fig. 7a) is attributed to the decomposition of hydrotalcite-like phases, which are commonly reported secondary products in alkali-activated slag systems[35,75,76,102]. The formation of hydrotalcite-like phases is confirmed by characteristic XRD reflection at 2θ ≈ 11.6°, 23.3°, and 34.9° (Fig. 9)[75,76]. In the FTIR spectra (Fig. 8a), these phases contribute to the H–O–H bending band near 1640–1650 $cm^{-1}$ and to bands in the 1400–1500 $cm^{-1}$ region associated with asymmetric C–O stretching vibrations of interlayer carbonate anions[85]. At higher temperatures (~550-800 °C), additional mass loss is attributed to the decomposition of calcium carbonate, originating either from trace carbonate phases in the raw GGBS or from mild carbonation during sample handling and storage. This assignment is supported by the appearance of an FTIR band near 875 $cm^{-1}$, corresponding to C–O bending vibrations in calcite [85].

In contrast, the fly ash-based systems exhibit substantially lower mass loss at all curing ages (Fig. 7b and Fig. S2), indicating a significantly lower amount of chemically bound water and, thus, a lower overall degree of reaction. This observation is consistent with the much lower cumulative heat release measured by ICC (Fig. 6b). FTIR spectra of the activated FA samples (Fig. 8b) show only minor shifts of the main asymmetric Si–O–T stretching band toward lower wavenumbers over 28 days, suggesting limited gel formation. Similarly, the 28-day XRD patterns (Fig. 9) closely resemble those of the corresponding raw fly ash, with no discernible crystalline reaction products detected. This observation is consistent with the predominantly amorphous nature of the reaction product in alkali-activated fly ash, commonly denoted as N-A-S-(H) gel.

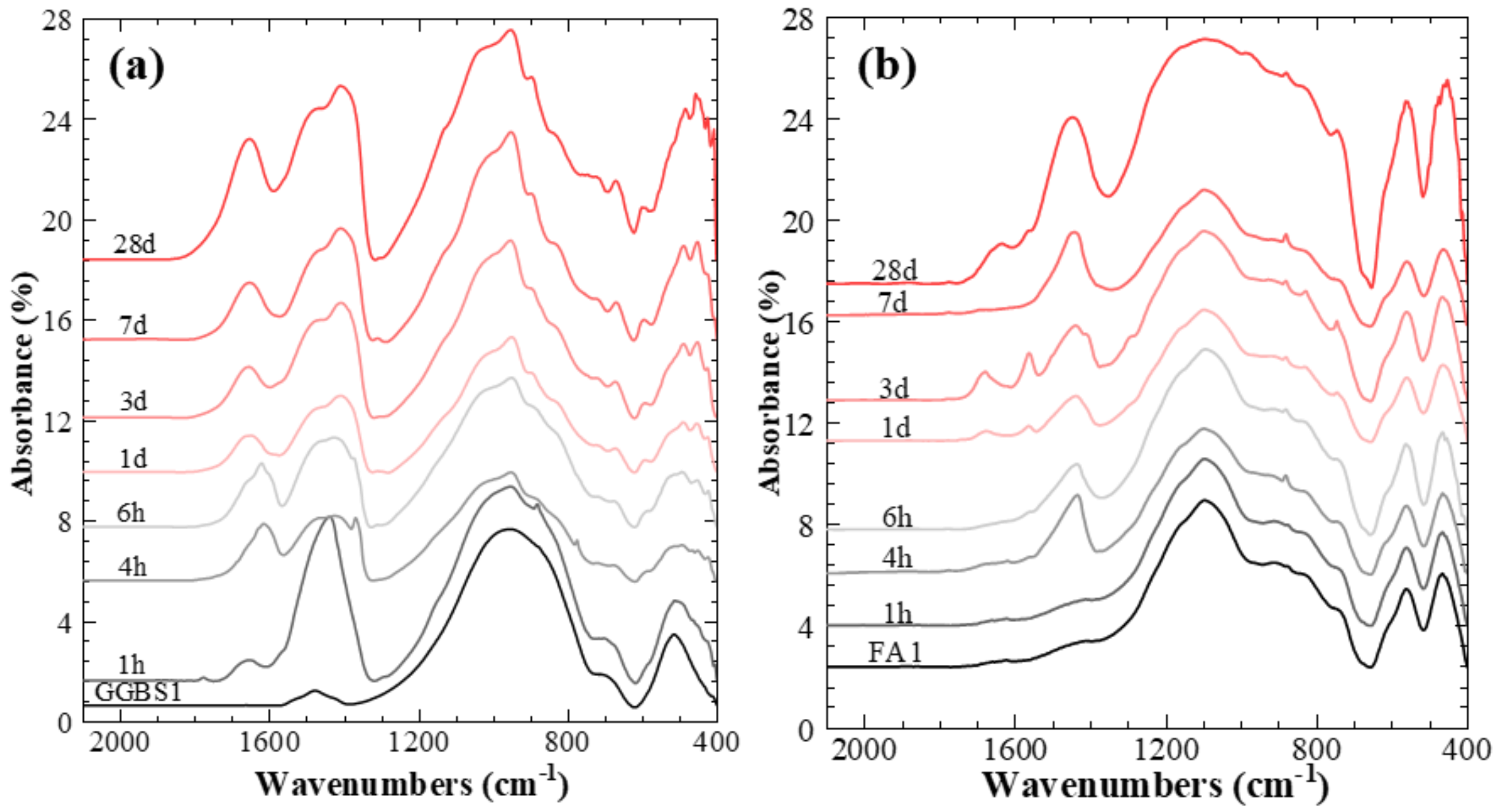

**Fig. 8.** Evolution of FTIR spectra in the range of 400-2000 $cm^{-1}$ for NaOH-activated (a) GGBS1, and (b) FA1 binders at different curing ages up to 28 days.

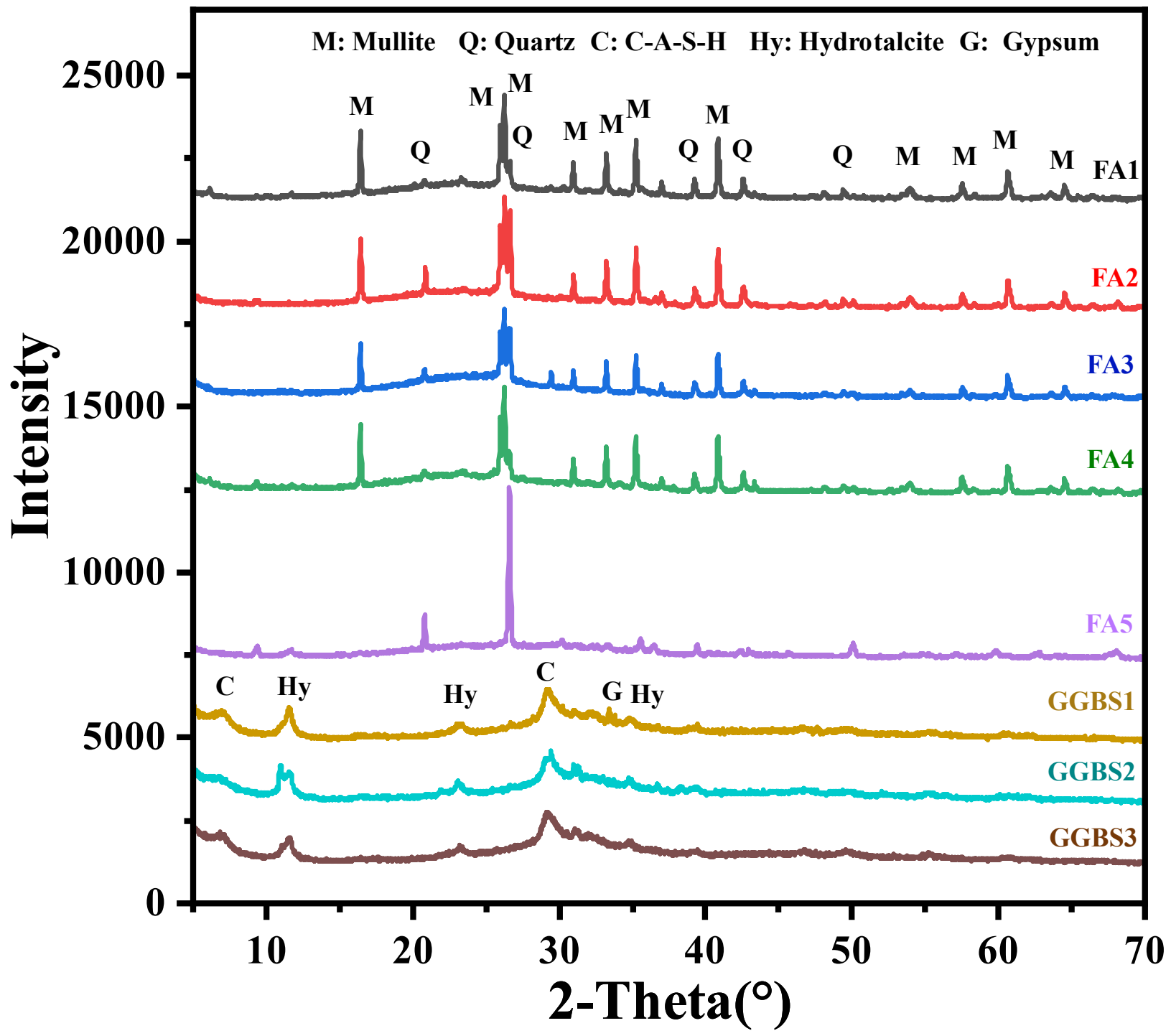

**Fig.9.** XRD patterns of NaOH-activated fly ash (FA1-FA5) and slag (GGBS1-GGBS3) samples after 28 days of curing.

To quantitatively assess these trends, the bound water contents of all samples at 3, 7, and 28 days were calculated from the TGA mass loss between 30 °C and 650 °C, following similar protocols in refs. [75,76]. Contributions from the raw precursors were subtracted to isolate the bound water associated with reaction products. The resulting bound water contents are presented in Fig. 10a and show a progressive increase with curing age across all samples, reflecting continued reaction and accumulation of gel phases over time. Slag-based binders consistently exhibit higher bound water contents than fly ash-based binders under identical activation conditions, reflecting their higher reactivity. Within the fly ash group, the high-Ca FA5 sample exhibits the highest bound water content, whereas FA3 shows the lowest, indicating the lowest reactivity among the fly ashes examined. Moreover, a strong linear correlation between TGA bound water content and ICC cumulative heat release (Fig. 10b) demonstrates that these metrics provide consistent and complementary measures of reaction progress.

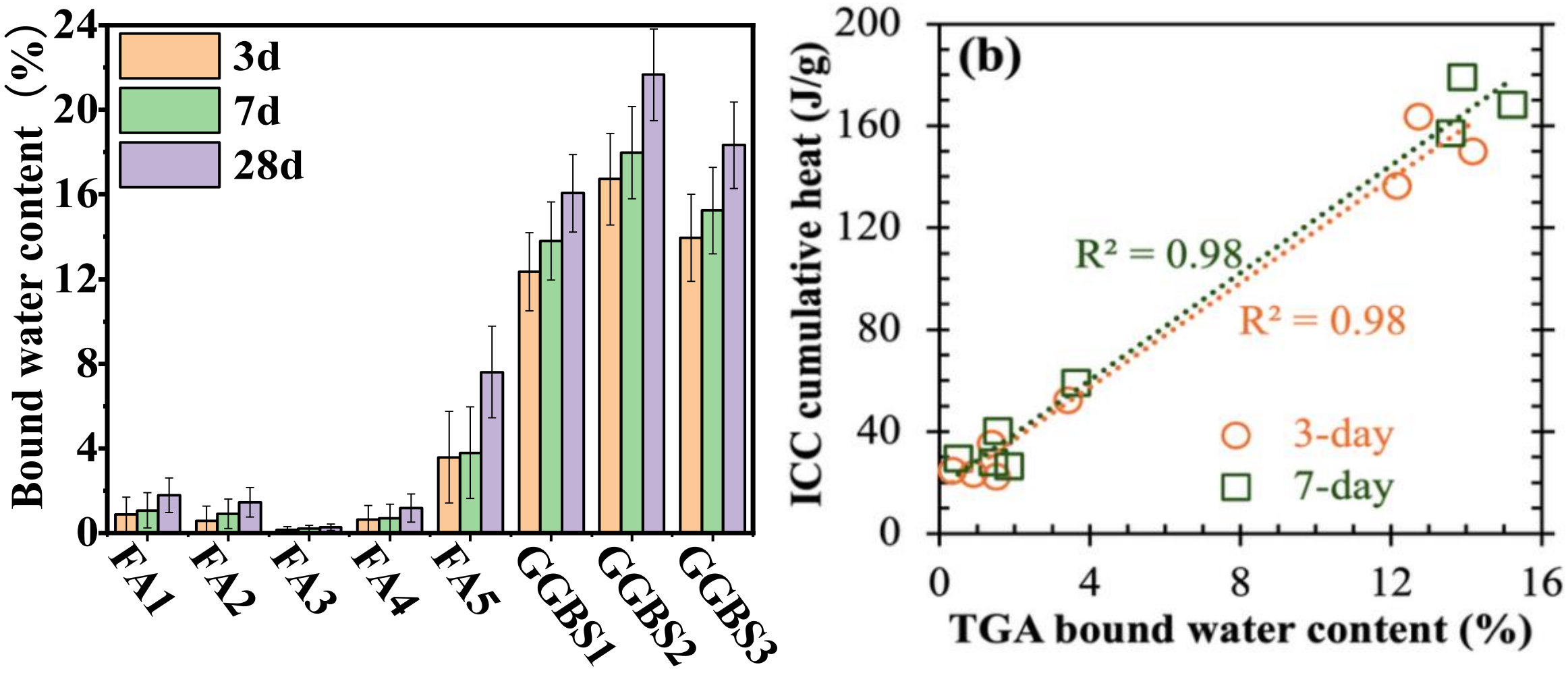

Fig. 10. Bound water content measured by TGA for NaOH-activated fly ash (FA1-FA5) and slag (GGBS1-GGBS3) samples at 3, 7 and 28 days of curing. (b) Correlation between TGA bound water content (at 3 and 7 days) and ICC cumulative heat release.

### *4.3.3 Compressive strength*

The compressive strengths of all mixes at 1, 3, 7 and 28 days are summarized in Table 9 and shown in Fig. 11. Each reported value represents the mean of at least three 50-mm cubic paste specimens. Alkali-activated slag binders (GGBS1-GGBS3) exhibits markedly higher strengths with rapid early-age development, achieving 1-day strengths of 19.0–24.6 MPa, followed by moderate gains to ~23.6–26.2 MPa at 28 days. This rapid strength development is consistent with the pronounced early exothermic behavior observed in ICC measurements (Fig. 6) and with TGA and FTIR evidence of rapid formation of C-(N)-A-S-H-type gel (Figs. 7 and 8). In contrast, alkali-activated fly ash binders exhibit significantly lower strength development under identical activation conditions (3 M NaOH). After 28 days of curing, FA1–FA4 reach compressive strengths of only 1.0–2.4 MPa, whereas FA5 reaches approximately 13 MPa. The limited strength development of FA1–FA4 correlates with their low reactivity, as indicated by ICC results (Fig. 6) and TGA bound water contents (Fig. 10a). The compressive strengths reported here should be interpreted as comparative outcomes obtained under a uniform activator formulation, rather than as optimized values for fly ash-based systems. Given the intrinsically lower reactivity of most fly ashes relative to slag, higher alkali concentrations or alternative activation strategies are typically required to achieve substantially improved strengths [48].

Table 9. Summary of compressive strength (in MPa) at 1, 3, 7, and 28 days for NaOH-activated fly ash (FA1–FA5) and slag (GGBS1–GGBS3) binders. Values are reported as the means of at least three replicates.

| | FA1 | FA2 | FA3 | FA4 | FA5 | GGBS1 | GGBS2 | GGBS3 |
|---|---|---|---|---|---|---|---|---|
| 1d | 0.68 | 0.27 | 0.38 | 0.38 | 4.59 | 19.03 | 24.61 | 21.94 |
| 3d | 0.79 | 0.37 | 0.40 | 0.59 | 6.73 | 20.11 | 25.38 | 22.88 |
| 7d | 0.97 | 0.45 | 0.65 | 0.77 | 6.89 | 21.33 | 26.20 | 22.88 |
| 28d | 2.4 | 1 | 2.2 | 2.2 | 13 | 25 | 26.2 | 23.6 |

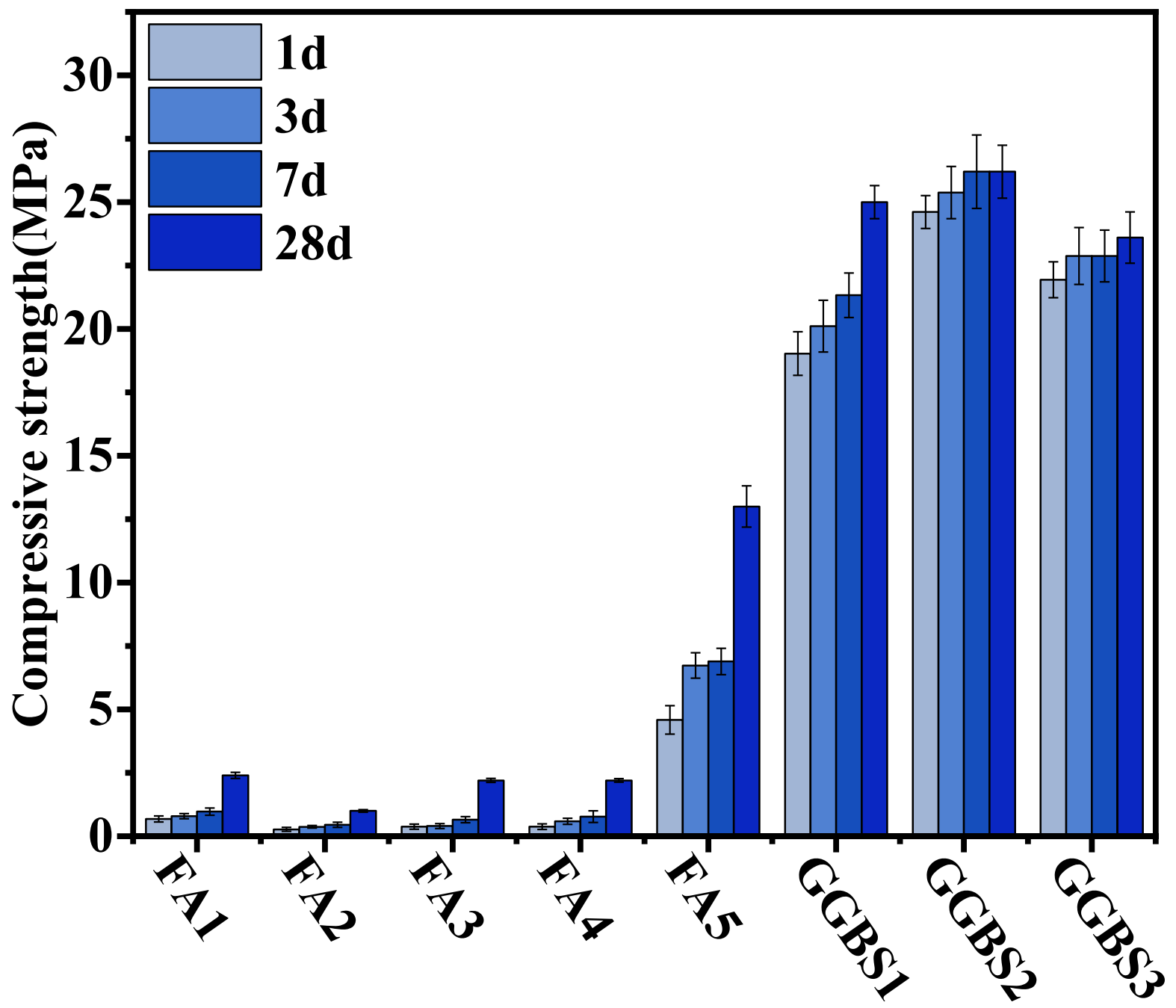

Fig. 11. Compressive strength development of NaOH-activated fly ash (FA1–FA5) and slag (GGBS1–GGBS3) binders at 1, 3, 7, and 28 days of curing. Values represent the mean of ≥ 3 50-mm cube specimens, with error bars corresponding to one standard deviation.

### 4.4 Performance of Structural Descriptors in Capturing Reactivity

This section evaluates the performance of the structural descriptors introduced in Section 3.4, namely, AMOBS and AMODE, along with their modified forms, to capture intrinsic reactivity trends of the FA and GGBS precursors. Given the strong linear correlations observed between AMOBS and AMODE and their modified forms (Fig. 5), as well as strong consistency among the experimental reactivity indicators presented in Section 4.3, the analysis here focuses on assessing whether the modified AMODE parameter (Equation (6)) can effectively capture reactivity trends derived from ICC data. For completeness, additional comparisons involving other three descriptors and alternative reactivity metrics (TGA bound water content and compressive strength) are provided in Figs. S4-S14 of the Supplementary Material and show trends consistent with those discussed here.

The initial ICC peak is dominated by surface wetting and rapid initial physicochemical interactions, which are observed even in low-reactivity fly ashes and therefore do not reliably reflect intrinsic precursor reactivity governed by glass-network bond rupture. Accordingly, the first 0.5 h of heat release was excluded to isolate the contribution from alkaline activation reactions. The cumulative reaction heat ($H_t - H_{0.5}$) was further normalized by both BET surface area, as in previous studies[31,100], and by amorphous content, since only the amorphous fraction actively participates in dissolution and gel formation [80]. The normalization details are provided in Section S8 of the Supplementary Material. The normalized ICC cumulative heats of the five FA and three GGBS samples at different ages are plotted against their corresponding modified AMODE values in Fig. 12a. A strong inverse relationship is observed ($R^2$ = 0.99–1.00 for exponential fits): higher AMODE values correspond to lower normalized reaction heats. This behavior is consistent with the physical interpretation of AMODE as a measure of average energy required to rupture metal–oxygen bonds within the amorphous network: precursors with higher AMODE values possess stronger metal–oxygen bonding and are therefore less susceptible to dissolution under alkaline conditions, resulting in reduced reaction heat release.

To extend the compositional range beyond the relatively narrow variations within the low-Ca fly ashes (FA1–FA4) and slags (GGBS1–GGBS3), additional ICC tests were performed on four blended precursor systems: (1) 70%GGBS3 + 30%FA5, (2) 30%GGBS3 + 70%FA5, (3) 70%FA1 + 30%FA5, and (4) 30% FA1 + 70%FA5. Complete ICC curves for these blends, collected under identical activation conditions, are provided in Fig. S15 of the Supplementary Material. For each blend, normalized reaction heats were calculated using weighted BET surface areas and amorphous contents, while the corresponding modified AMODE values were determined as weighted averages of the individual precursor values. The combined dataset comprising both single-precursor and blended systems (Fig. 12b) confirms that the inverse relationship between modified AMODE and normalized ICC cumulative heat is preserved across this broader compositional space, with $R^2$ values of 0.88–0.93 for exponential fits. Together, these results demonstrate that the modified AMODE parameter, along with other three descriptors examined (Section S9 of the Supplementary Material), provides a robust and physically interpretable metric for capturing relative intrinsic reactivity trends across a wide range of precursor compositions, including both single and binary systems.

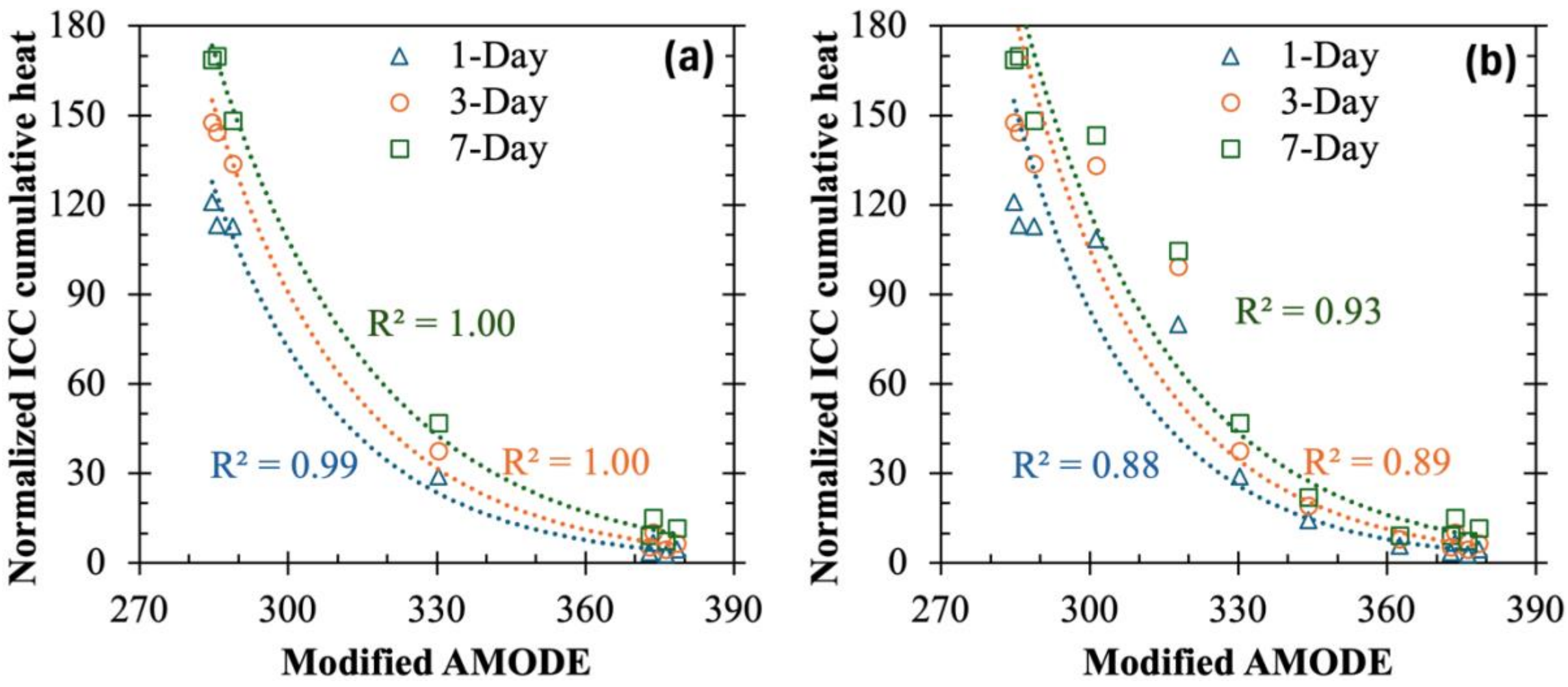

Fig. 12. Relationship between modified AMODE values (calculated using Equation (6)) and normalized ICC cumulative heat release for NaOH-activated systems: (a) single precursors (five FAs and three GGBSs), and (b) combined dataset including both single-precursor and blended systems.

## 4.5 Broader impact and limitations

### *4.5.1 Broader impact*

This study demonstrates that physics-based reactivity descriptors originally developed for fully amorphous systems can be systematically extended to heterogeneous, real-world precursors containing both amorphous and crystalline phases, including coal fly ashes, GGBSs, and their binary blends. By combining quantitative phase analysis (via Rietveld-based QXRD) with bulk chemical composition obtained from XRF, the effective composition of the amorphous fraction was isolated and used to generate atomistic structural representations. This framework enables the extension of structural descriptors such as the AMODE and AMOBS and their modified versions to a broader range of chemical and mineralogical compositions than previously examined.

Strong and consistent correlations are observed between these descriptors and multiple independent experimental indicators of reactivity, including ICC cumulative heat release, TGA bound water content, and compressive strength. The compositional space spanned by the evaluated descriptors includes multicomponent aluminosilicate systems containing $SiO_2$, $Al_2O_3$, $TiO_2$, $Fe_2O_3$, CaO, MgO, MnO, $Na_2O$, and $K_2O$, which represent the major oxides in common SCMs and AAM precursors. From a practical perspective, the proposed framework provides a transferable, structure-informed basis for comparative assessment of intrinsic precursor reactivity. This capability is particularly valuable for screening chemically and mineralogically variable industrial byproducts and for informing precursor selection and blending considerations, while complementing experimental testing. By accounting for mineralogical heterogeneity and amorphous-phase chemistry, the framework moves beyond empirical composition-only indicators and enables a physically grounded evaluation of reactivity trends across diverse precursor sources.

More broadly, the approach is conceptually general and may be extended to other amorphous–crystalline precursor families, including natural pozzolans, calcined clays, waste glass cullet blends, brick powder, certain mine tailings, and red mud-derived materials, provided that the amorphous fraction can be reliably quantified and mapped into oxide composition space. Beyond cementitious materials, similar descriptor-based strategies may be transferable to other mixed amorphous–crystalline systems in which dissolution-controlled reactivity or chemical durability is governed by atomic-scale bond energetics, including materials used in waste stabilization, mineral carbonation, and other geochemical or environmental reaction systems.

#### *4.5.2 Limitations and future work*

While this study extends physics-based glass-structure reactivity descriptors to mixed amorphous–crystalline precursor systems, several limitations should be acknowledged and considered during applications. First, as with prior structural descriptors and composition-based parameters, the present approach implicitly assumes congruent dissolution of oxide components within the amorphous fraction. In practice, dissolution behavior of amorphous aluminosilicate can be incongruent and strongly dependent on glass composition and environmental conditions, including solution chemistry, pH, and temperature[105–108]. Incongruent dissolution may lead to the formation of altered surface layers or secondary reaction products that retard subsequent dissolution[109,110]. Future work should therefore evaluate the impact of such effects and where significant, explore how descriptors such as modified AMODE and AMOBS might be adapted to account for non-congruent behavior, for example through surface-controlled terms or condition-dependent weighting schemes.

Second, although the present study evaluates precursor reactivity under NaOH-based activation, the structural descriptors examined here reflect intrinsic properties of the amorphous aluminosilicate network, particularly the energetic stability of metal–oxygen bonds. As such, the relative reactivity trends captured by modified AMODE and AMOBS are expected to be transferable to other alkaline environments in which dissolution of the amorphous phase governs reaction progress. Nevertheless, activator chemistry (e.g., hydroxide versus silicate-based systems, sodium versus potassium activators), curing temperature, and liquid-to-solid ratio can influence reaction kinetics, phase assemblage, and absolute reaction extents. Consequently, while general inverse trends with descriptor values are expected to persist, quantitative correlations may require protocol-specific calibration under different activation or curing regimes.

Third, the analysis evaluates descriptors exclusively on the amorphous fraction, treating crystalline components as inert. This assumption is appropriate for the dominant crystalline phases identified in this study (e.g., quartz and mullite), which exhibit minimal reactivity under the alkaline conditions employed. However, caution is warranted when extending the framework to materials containing more reactive crystalline phases (e.g., calcium silicates or zeolitic minerals), which may contribute non-negligibly to reaction under certain alkaline environments and would require explicit consideration.

Fourth, the amorphous fraction is treated as a single chemically homogeneous phase. Prior studies of fly ashes have reported chemically distinct glass populations within the amorphous fraction[111,112]. Such phase-level heterogeneity can influence experimental reactivity measurements and is not explicitly resolved in the present modeling framework. Accordingly, the descriptors developed here, as well as those in related studies[30–34], should be interpreted as effective, averaged measures of amorphous network energetics. Incorporating multi-glass representations or population-weighted descriptors represents an important direction for future refinement.

Finally, there are uncertainties associated with QXRD-based amorphous quantification[113,114], XRF compositional analysis[115], and BET surface area measurements[41,116], which propagate into the calculated descriptor values. These uncertainties should be considered when interpreting prediction accuracy, particularly when the intrinsic reactivity differences among precursors are small. At the structural level, the descriptors rely on effective averages of coordination environments and M–O bond energetics and do not explicitly resolve coordination-dependent energetics, mixed valence states, specific local structural motifs or the evolution of altered surface layers during reaction. Moreover, the average coordination numbers used in calculating modified AMODE and AMOBS are derived from datasets spanning a finite compositional range, particularly for minor oxides, and may require refinement as broader data become available. These simplifications facilitate practical application but may limit mechanistic fidelity. Despite these limitations, within the compositional space examined in this study, the strong correlations observed across different precursors and their blends demonstrate the utility of the proposed descriptors for relative reactivity screening for formulating cementitious materials.

## 5 Conclusions

This study demonstrates that the intrinsic reactivity of heterogeneous industrial by-products such as fly ash (FA), ground granulated blast-furnace slag (GGBS) and their FA-GGBS blends can be systematically assessed using physics-informed glass-structure descriptors, even in the presence of mixed amorphous and crystalline phases. While such descriptors were originally developed and validated primarily for ideal or predominantly amorphous systems, their application to real industrial precursors with mixed amorphous-crystalline assemblages has been constrained by the need of isolating amorphous-phase composition in mineralogically complex materials.

To address this limitation, quantitative X-ray diffraction combined with bulk chemical analysis was used to reconstruct the chemical compositions of the amorphous fractions in five fly ashes and three GGBSs. Molecular dynamics simulations employing a melt-and-quench approach were then used to generate atomic-scale structural models of the corresponding glassy phases. Based on these structures, previously introduced descriptors, namely average metal oxygen dissociation energy (AMODE) and average metal oxygen bond strength (AMOBS), along with their modified, composition-based forms, were refined to cover a broad compositional space including $SiO_2$, $Al_2O_3$, $TiO_2$, $Fe_2O_3$, CaO, MgO, MnO, $Na_2O$, and $K_2O$.

The refined descriptors exhibit strong inverse correlations with multiple independent reactivity indicators, including cumulative heat release from isothermal calorimetry, nonevaporable water content from thermogravimetric analysis, and compressive strength, for both single precursors and binary FA–GGBS blends activated with NaOH. Higher AMODE and AMOBS values, which reflect stronger average metal-oxygen bonding within the glass network, correspond to lower reactivity across these metrics. Because these descriptors are derived directly from glass-network bond energetics, they capture a material-intrinsic component of precursor reactivity that is largely insensitive to the specific activation protocol. Nevertheless, experimentally measured reactivity remains influenced by solution chemistry, which governs reaction kinetics and overall reaction extent.

More broadly, this work establishes a transferable, material-intrinsic framework linking composition, atomic-scale structure, and relative reactivity in precursors containing mixed amorphous and crystalline phases. By integrating quantitative phase characterization with physics-informed structural descriptors, the proposed framework enables systematic, structure-informed comparative assessment of precursor reactivity. This approach complements experimental testing and offers quantitative insight that may inform precursor selection and blending considerations in cementitious material systems.

## 6 Acknowledgement

Z.P., X.L., and Y.W. acknowledge support from the National Natural Science Foundation of China under Grant Nos. 52293434 and U20A20313. K.G. and S.H. acknowledge support from the Department of Civil and Environmental Engineering at Rice University.

## 7 Supplementary Material

The crystallographic information files (CIF) used for phase identification; The potential parameters of the Pedone force field; Calculation of radial distribution function and coordination number; Rietveld refinement of XRD patterns; Average metal-oxygen bond strength (AMOBS) parameter; Additional TGA data; Full FTIR spectra; Normalization of reactivity data; Additional comparisons involving other three descriptors and ICC data; Comparisons of the four descriptors with TGA bound water content data; Comparisons of the four descriptors with compressive strength data; ICC data for binary systems

## 8 Use of AI-Assisted Technologies

The authors used ChatGPT(5.2) to assist with language refinement during manuscript preparation. All scientific content, interpretations, and conclusions were developed and verified by the authors, who reviewed and revised the text as needed and take full responsibility for the final submitted manuscript.

## 9 Declaration of Competing Interest

The authors declare that they have no known competing financial interests or personal relationships that could have appeared to influence the work reported in this paper.